\documentclass[prb,aps,twocolumn,amsmath,amssymb,floatfix,showpacs,
superscriptaddress]{revtex4}

\usepackage[dvips]{graphics}
\usepackage{color}
\definecolor{dblue}{rgb}{0,0,0.6}

\begin{document}

\title{\textcolor{dblue}{Quantum transport in mesoscopic ring structures: 
Effects of impurities, long-range hopping and interactions}}

\author{Santanu K. Maiti}

\email{santanu.maiti@isical.ac.in}

\affiliation{Physics and Applied Mathematics Unit, Indian Statistical
Institute, 203 Barrackpore Trunk Road, Kolkata-700 108, India}

\begin{abstract}

In the present review we make a comprehensive analysis of our understanding
on electron transport in mesoscopic single-channel rings and multi-channel
cylinders within a tight-binding framework. A spectacular mesoscopic 
phenomenon where a non-decaying current circulates in a small conducting 
loop is observed upon the application of an Aharonov-Bohm flux $\phi$.
To understand its behavior one has to focus attention on the interplay
of quantum phase coherence, electron-electron correlation and disorder.
This is a highly challenging problem and here we address it for some
simple loop geometries with their detailed energy band structures to get 
an entire picture at the microscopic level. The behavior of low-field
magnetic response of persistent current and its temperature dependence
are also discussed.

\end{abstract}


\maketitle 

\section{Introduction}

An emerging tendency in modern material science is to propose and investigate
systems containing smaller and smaller structures. These smaller structures
approach the so-called mesoscopic or nanoscopic regimes in which quantum
effects become much more significant for the behavior of these materials. 
This situates mesoscopic physics at the interface of statistical and
quantum pictures. The mesoscopic systems are very much smaller than the 
large-scale objects and they often have unusual physical and chemical 
properties. The study of such systems provides a clear understanding of 
the behavior of a material as it goes from a few atoms to large visible 
and tangible objects.

\subsection{The mesoscopic regime}

The mesoscopic scale refers to the length scale at which one can 
reasonably describe the properties of a material or a phenomenon without 
discussing the behavior of the individual atoms. For solids it is 
typically a few to ten nanometers and involves averaging over a few 
thousand atoms or molecules. In this scale the expected fluctuations 
of the averaged physical quantities due to the motion and behavior of 
individual particles can be reduced below some desirable threshold 
(often to a few percent) and it must be rigorously established
within the context of any particular problem. In the mesoscopic regime, 
behavior of a system is considerably influenced by quantum interference
of electronic wave functions. The quantum phase coherence, essential 
for the appearance of interference effects, is preserved only during a 
finite time $\tau_{\phi}$ the so-called phase coherence time. In electronic 
conductors, finite phase coherence time corresponds to a phase coherence 
length $L_{\phi}$ over which electrons can travel before their phase 
coherence gets lost. Mesoscopic quantum effects appear when the typical 
time or length scales of the system are smaller than the phase coherence 
time or length. In many cases this means that the relevant system size $L$ 
must be smaller than the phase coherence length $L_{\phi}$. For an electron, 
the phase coherence time/length is limited by electron-electron and 
electron-phonon scattering. These processes are important at high 
temperatures, but both are suppressed at low temperatures implying that the 
phase coherence time/length is strongly material and temperature dependent.

The mesoscopic regime is therefore characterized by small time and/or 
length scales and low temperatures. When temperature is lowered, the phase
coherence time/length increases (by a factor $T^{-1}$), and the mesoscopic
regime gets extended. At sub-Kelvin temperatures, the time and the length 
scales in semiconductor samples are of the order of picoseconds and 
micrometers, respectively. 

\subsection{Some extraordinary mesoscopic phenomena}

The samples like quantum dots, quantum wires, two-dimensional electron 
gases in semiconductor heterostructures, etc., exhibit many alluring 
physical properties. Here we briefly describe some spectacular effects 
that appear in such systems as a consequence of quantum phase coherence 
of electronic wave functions.

\subsubsection{Aharonov-Bohm oscillation}

One of the most notable consequences of quantum phase coherence is the
Aharonov-Bohm (AB) oscillation in conductance of normal metal mesoscopic 
rings. At low temperatures superposition of electronic waves propagating 
along two arms of a ring becomes important. The pioneering experiment on 
AB effect was done on a ring-shaped resistor made from a $38$ nm film of 
polycrystalline gold. The diameter of it is $820$ nm and the thickness 
of the wire is $40$ nm~\cite{webb1}. The conductance shows oscillating
behavior as a function of magnetic flux enclosed by the ring with $h/|e|$ 
periodicity~\cite{webb2}:
\begin{equation}
g=g_0 + \hat{g} \cos\left[\frac{|e|BS}{\hbar} + \overline{\phi} \right] 
\end{equation}
where $S$ is the area enclosed by the ring and $B$ is the applied magnetic 
field perpendicular to the ring plane.

\subsubsection{Integer quantum Hall effect}

One of the most stunning discoveries of $1980$s was the integer quantum
Hall effect (IQHE)~\cite{klit} as a result of quantum phase coherence of 
electronic wave functions in two-dimensional electron gas ($2$DEG) systems. 
In the Hall measurement, one drives a current along a conductor ($2$DEG), 
and measures the longitudinal voltage $V_x$ and transverse Hall voltage 
$V_H$ as a function of magnetic field $B$ applied perpendicular 
to the plane of the conductor. According to the classical Drude formula, 
the Hall resistance $R_H$ varies linearly with the field strength $B$,
while the longitudinal resistance $R_x$ gets unaffected by this field. 
This behavior holds true only when magnetic field is too weak. In strong 
magnetic field and at low temperatures, one gets completely different 
behavior which cannot be explained by the classical Drude model. At high 
field longitudinal resistance shows oscillatory behavior, while Hall 
resistance exhibits step-like nature with sharp plateaus. The values of 
$R_H$ on these plateaus are given by $h/n e^2$, where $n$ is an integer 
with values $1,2,3, \ldots$ and it turns out that these values of $R_H$ 
are highly reproducible with a great precision and are also very robust so 
that they are often used as resistance standard. The integer quantum Hall 
effect is a purely quantum mechanical phenomenon due to the formation 
of the Landau levels and many good reviews on IQHE are available in 
literature~\cite{pran,chakra,imry}.

\subsubsection{Fractional quantum Hall effect}

At extremely high magnetic fields and low temperatures, a $2$DEG provides 
additional plateaus in the Hall resistance at fractional filling factors 
and this phenomenon was discovered in $1982$~\cite{tsui}. Unlike the integer 
quantum Hall effect, it has been observed that the Coulomb correlation 
between the electrons becomes important for the interpretation of the 
fractional quantum Hall effect and the presence of fractional filling has 
been traced back to the existence of correlated collective quasi-particle
excitations~\cite{laug}. This phenomenon has been reviewed in detail by 
Chakraborty {\em et al.}~\cite{chakra}.

\subsubsection{Conductance fluctuations and quantization}

In the mesoscopic regime, conductance of disordered wires exhibits pronounced 
fluctuations as a function of external parameters, like magnetic field or 
Fermi energy. These fluctuations were observed~\cite{fowl} only at very 
low temperatures which are perfectly reproducible and represent a fingerprint 
of the quantum effects in samples. The fluctuations appear due to the 
interference of electronic wave functions corresponding to different pathways 
that the electrons can take when traversing through a system. The most 
significant feature of this conductance fluctuations is that their typical 
amplitude is universal in the diffusive regime~\cite{lee}. The fluctuations 
are always of the order of the conductance quantum $e^2/h$ and depend only 
on the basic symmetries of the system~\cite{mail}.

The conductance of ballistic quantum point contacts was found~\cite{wees,wha} 
to be quantized in units of $2 e^2/h$ and a recent experiment~\cite{costa} 
demonstrates that the conductance quantization can be observed even in an 
extremely simple setup. A quantum point contact is a very narrow link 
between two conducting materials formed by imposing a narrow constriction 
into them. With the decrease of the width ($W$) of constriction it has been 
observed that conductance goes down in quantized steps. This is due to the 
fact that although the width of the constriction changes continuously, the 
number of sub-bands or transverse modes ($M$) changes in discrete steps. 
This discreteness is not evident if the constriction is several thousands 
of wavelengths wide, since then a very small fractional change in $W$ 
changes $M$ by many integers.

\subsubsection{Persistent current}

The physics of small conducting rings provides an excellent testing ground 
for many ideas of fundamental physics. In thermodynamic equilibrium, a small 
metallic ring threaded by an AB flux $\phi$ supports a current that 
does not decay dissipatively even at non-zero temperature. It is the 
well-known phenomenon of persistent current in mesoscopic normal metal 
rings. This is a purely quantum mechanical effect and gives a lucid 
exposition of the AB effect~\cite{aharo}. The possibility of persistent 
current was predicted in the very early days of quantum mechanics by 
Hund~\cite{hund}, but their experimental evidences came much later only 
after realization of the mesoscopic systems. In $1983$, B\"{uttiker} 
{\em et al.}~\cite{butt} predicted theoretically that persistent current 
can exist in mesoscopic normal metal rings threaded by a magnetic flux 
even in the presence of impurity. This quantum phenomenon has been verified
experimentally in a pioneering work of Levy {\em et al.}~\cite{levy} and 
later the existence of persistent current has been further confirmed 
by several other nice experiments~\cite{deb,reu,jari,chand,mailly1,blu,fvo}. 
Though much efforts have been paid to study persistent current both
experimentally~\cite{levy,deb,reu,jari,chand,mailly1,blu,fvo} as well as 
theoretically~\cite{butt1,cheu1,cheu2,mont,mont1,land,byers,von,bouc,alts,
schm,abra,mull,sm1,sm2,sm3,sm4,sm5,sm6,sm7,sm8,pc1,pc2,pc3,pc4,pc5}, yet 
several anomalies still exist between the theory and 
experiment, and the full knowledge about it in this scale is not well 
rooted even today. One of the important controversies is related to the 
concrete prediction of persistent current amplitudes in mesoscopic rings. 
The results of the single loop experiments are somewhat different from those 
for the ensemble of isolated loops. 
Levy {\em et al.}~\cite{levy} found
oscillations with period $\phi_0/2$ rather than $\phi_0$ in an ensemble
of $10^7$ independent Cu rings. Similar $\phi_0/2$ oscillations were also
reported for an ensemble of disconnected $10^5$ Ag rings~\cite{deb} as well
as for an array of $10^5$ isolated GaAs-AlGaAs rings~\cite{reu}. In a
recent experiment, Jariwala {\em et al.}~\cite{jari} obtained both $\phi_0$
and $\phi_0/2$ periodic persistent currents in an array of thirty diffusive
mesoscopic Au rings. Persistent currents with expected $\phi_0$ periodicity 
have been observed in isolated single Au rings~\cite{chand} and in a 
GaAs-AlGaAs ring~\cite{mailly1}. Except for the case of nearly ballistic 
GaAs-AlGaAs ring~\cite{mailly1}, all the measured currents are in general 
one or two orders of magnitude larger than those expected from the theory.

Free electron theory predicts that at absolute zero temperature ($T=0\,$K),
an ordered one-dimensional ($1$D) metallic ring threaded by a magnetic flux
$\phi$ supports persistent current with maximum amplitude $I_0=ev_F/L$, where
$v_F$ is the Fermi velocity and $L$ is the circumference of the ring.
Metals are intrinsically disordered which tends to decrease persistent
current, and the calculations show that the disorder-averaged current
$\langle I \rangle$ crucially depends on the choice of the
ensemble~\cite{cheu2,mont,mont1}. The magnitude of the current
$\langle I^2\rangle^{1/2}$ is however insensitive to averaging issues,
and is of the order of $I_0 l/L$, $l$ being the elastic mean free path
of electrons. This expression remains valid even if one takes into
account the finite width of the ring by adding contributions from 
transverse channels, since disorder leads to a compensation between the
channels~\cite{cheu2,mont}. However, the measurements on an ensemble
of $10^7$ Cu rings~\cite{levy} reported a diamagnetic persistent current
of average amplitude $3 \times 10^{-3}$ $ev_F/L$ with half a flux-quantum
periodicity. Such $\phi_0/2$ oscillations with diamagnetic response were
also found in other persistent current experiments consisting of ensemble
of isolated rings~\cite{deb,reu}.

Measurements on single isolated mesoscopic rings on the other hand detected 
$\phi_0$-periodic persistent currents with amplitudes of the order of 
$I_0 \sim ev_F/L$, (closed to the value for an ordered ring). Theory and 
experiment~\cite{mailly1} seem to agree only when {\em disorder is weak}. 
In another recent experiment Bluhm {\em et al.}~\cite{blu} have measured 
magnetic response of $33$ individual cold mesoscopic gold rings, one ring 
at a time, using a scanning SQUID technique. They have measured $h/e$ 
component and anticipated that the measured current amplitude agrees quite 
well with theory~\cite{cheu1} in a single ballistic ring~\cite{mailly1} and 
an ensemble of $16$ nearly ballistic rings~\cite{raba}. However, the
amplitudes of the currents in single-isolated-diffusive gold 
rings~\cite{chand} were two orders of magnitude larger than the
theoretical estimates. This discrepancy instituted intense theoretical
activity, and it is generally believed that the electron-electron
correlation plays an important role in the disordered diffusive
rings~\cite{abra,bouz,giam}. An explanation based on the perturbative 
calculation in presence of interaction and disorder has been proposed and 
it seems to give a quantitative estimate closer to the experimental results, 
but still it is less than the measured currents by an order of magnitude, 
and the interaction parameter used in the theory is not well understood
physically.

The other paramount debate arises in determining the sign of low-field 
currents and still it is an unresolved issue between theoretical and 
experimental results. In an experiment Levy {\em et al.}~\cite{levy} have 
shown diamagnetic nature of the measured currents at low-field limit. 
Jariwala {\em et al.}~\cite{jari} have predicted diamagnetic persistent 
current in their experiment and similar diamagnetic response in the 
vicinity of zero-field limit were also supported in an experiment done by 
Deblock~\cite{deb} {\em et al.} on Ag rings. In other experiment Chandrasekhar 
{\em et al.}~\cite{chand} have obtained paramagnetic response near zero-field 
limit. Yu and Fowler~\cite{yu} have 
shown both diamagnetic and paramagnetic responses in mesoscopic Hubbard 
rings. Though in a theoretical work Cheung {\em et al.}~\cite{cheu2} have 
revealed that the direction of current is random depending on total number 
of electrons in the system and the specific realization of random potentials. 
Hence, prediction of the sign of low-field currents is still an open 
challenge and further studies on persistent current in mesoscopic systems 
are needed to remove the existing controversies.

The main motivation of the present review is to address some important 
aspects of persistent current and low-field magnetic susceptibility in 
single-channel mesoscopic rings and multi-channel cylinders which are 
quite challenging from the standpoint of theoretical as well as experimental 
research. A brief layout of the presentation is as follows.

Following a short description of some extraordinary mesoscopic phenomena
(Section I), in Section II we study the phenomenon of persistent current 
in ordered and disordered rings within a non-interacting electron picture. 
The effects of electron-electron interaction on persistent current are 
explored in Section III. In Section IV, we show that the contributions 
of higher order hopping integrals are quite important for the enhancement 
of persistent current in disordered mesoscopic rings. The sign of persistent
current in the limit $\phi \rightarrow 0$, determined by calculating magnetic 
susceptibility, and its temperature dependence are analyzed in Section V. 
Finally, we conclude in Section VI.

Throughout the review we choose $c=e=h=1$ for numerical calculations.

\section{Persistent current in non-interacting single-channel and 
multi-channel mesoscopic rings}

Our aim of this section is to inspect the behavior of persistent current 
in some small non-superconducting loops threaded by a magnetic flux $\phi$. 
A conducting ring, penetrated by a magnetic flux $\phi$, carries an 
equilibrium current in its ground state that {\em persists} (does not decay) 
in time. An electrically charged particle moving around the ring but not 
entering the region of magnetic flux, feels no (classical) force during its 
motion. However, the magnetic vector potential $\vec{A}$, associated with 
magnetic field $\vec{B}$ through the relation 
$\vec{B}=\vec{\bigtriangledown} \times \vec{A}$, affects the quantum state 
of the particle by changing the phase of its wave function. As a consequence, 
both thermodynamic and kinetic properties oscillate with flux $\phi$. Here, 
we present some analytical and numerical results to study the characteristic 
properties of persistent current $I$ in mesoscopic rings as functions of AB 
flux $\phi$, system size $L$, total number of electrons $N_e$, chemical 
potential $\mu$ and disorder strength $W$.

\subsection{General formulation of persistent current}

In this sub-section we illustrate the theoretical formulation of persistent
current appears in a metallic ring in presence of an AB flux $\phi$. Let us
\begin{figure}[ht]
{\centering\resizebox*{5.25cm}{3cm}{\includegraphics{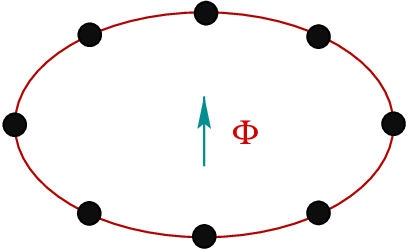}}\par}
\caption{(Color online). One-dimensional ring threaded by an AB flux $\phi$. 
Filled circles correspond to the positions of the atomic sites. A persistent 
current $I$ is established in the ring.}
\label{figure1}
\end{figure}
begin with the model quantum system given in Fig.~\ref{figure1}. The electric 
field $\mathcal E$ associated with magnetic field $B$ in the ring can be 
expressed through the relation (Faraday's law),
\begin{equation}
\vec{\bigtriangledown}\times \vec{\mathcal E}=-\frac{1}{c}\frac{\partial 
\vec{B}}{\partial t}.
\label{equ1}
\end{equation}
This relation enables us to determine the electric field $\mathcal E$ 
from the following expressions:
\begin{equation}
\oint_{S}(\vec{\bigtriangledown}\times \vec{\mathcal E}).\vec{dS}=
-\frac{1}{c}\oint_{S}\frac{\partial\vec{B}}{\partial t}.\vec{dS}=-\frac
{1}{c}\frac{\partial}{\partial t}\oint_{S}\vec{B}.\vec{dS}=-\frac{1}{c}
\frac{\partial\phi}{\partial t}
\label{equ2}
\end{equation}
where, $S$ is the area enclosed by the ring. From Stoke's theorem we
get,
\begin{equation}
\oint_{loop}\vec{\mathcal E}.\vec{dl}=-\frac{1}{c}\frac{\partial \phi}
{\partial t}
\label{equ3}
\end{equation}
and thus the electric field can be written in terms of the time 
variation of magnetic flux $\phi$ as,
\begin{equation}
\mathcal E = - \frac{1}{2\pi r c}\frac{\partial \phi}{\partial t}
\label{equ4}
\end{equation}
where $r$ is the radius of the ring. Therefore, the force experienced
by an electron in the ring becomes,
\begin{equation}
F=-\frac{e}{2\pi r c}\frac{\partial \phi}{\partial t}
\label{equ5}
\end{equation}
and the change in energy or work done for a small displacement
$\bigtriangleup s$ is,
\begin{equation}
\bigtriangleup E=\bigtriangleup W=\vec{F}.\vec{\bigtriangleup s}=
-\frac{e}{2\pi r c}\frac{\bigtriangleup \phi}{\bigtriangleup t}
\bigtriangleup s=-\frac{e}{2\pi r c}\bigtriangleup \phi\left(\frac
{\bigtriangleup s}{\bigtriangleup t}\right).
\label{equ6}
\end{equation}
The velocity of the electron in the ring is thus written as,
\begin{equation}
v = \frac{\bigtriangleup s}{\bigtriangleup t}= -\frac{2\pi r c}{e}
\left(\frac{\bigtriangleup E}{\bigtriangleup \phi} \right)
\label{equ7}
\end{equation}
and accordingly, the persistent current developed in the ring gets the
form,
\begin{equation}
I=ef=\frac{ev}{2\pi r}=-c\left(\frac{\bigtriangleup E}{\bigtriangleup \phi}
\right).
\label{equ8}
\end{equation}
This is the most general expression of persistent current in a loop geometry
threaded by an AB flux $\phi$ and it shows that the current can be obtained 
by taking the first order derivative of energy with respect to flux $\phi$.

\subsection{Non-interacting one-channel rings}

This sub-section focuses attention on the behavior of persistent current 
in one-channel rings~\cite{sanu1} where the ring are described by a simple 
non-interacting tight-binding (TB) framework. The TB Hamiltonian for a 
$N$-site ring threaded by a magnetic flux $\phi$ (measured in unit of the 
elementary flux quantum $\phi_0=ch/e$) reads,
\begin{equation}
H = \sum_{i}\epsilon_{i}c_i^{\dagger}c_i + \sum_{<ij>} t \left[e^{i\theta}
c_i^{\dagger} c_j + e^{-i\theta} c_j^{\dagger} c_i \right]
\label{equ9}
\end{equation}
where, $c_i^{\dagger}$ ($c_i$) corresponds to the creation (annihilation)
operator of an electron at the site $i$, $t$ gives the nearest-neighbor 
hopping integral, $\epsilon_i$ is the on-site energy and $\theta=2 \pi 
\phi/N$ is the phase factor due to the flux $\phi$ threaded by the ring. 
The magnetic flux $\phi$ enters explicitly into the above Hamiltonian 
(Eq.~\ref{equ9}), and the wave functions satisfy the periodic boundary 
condition which is equivalent to consider the above Hamiltonian at zero 
flux with the flux-modified boundary conditions:
\begin{eqnarray}
\left .\psi\right|_{x=L} & = & exp\left[\frac{2\pi i\phi}{\phi_{0}}\right]
\left .\psi\right|_{x=0} \nonumber \\
\left .\frac{d\psi}{dx}\right|_{x=L} & = & exp\left[\frac{2\pi i\phi}{\phi_0}
\right]\left .\frac{d\psi}{dx}\right|_{x=0}.
\label{equ10}
\end{eqnarray}
Here, $x$ varies between $0$ to $L$ and is expressed as 
$x=L\theta^{\prime}/2\pi$, where $\theta^{\prime}$ is the azimuthal angle, 
the spatial degrees of freedom of an electron in the ring. 

\subsubsection{Impurity free rings}

In order to reveal the basic properties of persistent current, let us 
start our discussion with the energy band structure of a simplest possible 
system which is the case of impurity free non-interacting electron model.

\begin{figure}[ht]
{\centering\resizebox*{7.75cm}{4.5cm}{\includegraphics{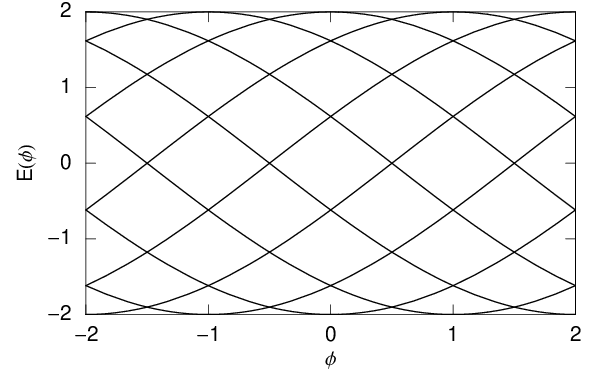}}\par}
\caption{Energy levels as a function of flux $\phi$ for a one-dimensional 
impurity-free ring when $N$ is set at $10$.}
\label{figure2}
\end{figure}

\vskip 0.2cm
\noindent
{\underline{Energy spectra}:} For a perfect ring, setting $\epsilon_i=0$
we get the energy of $n$th eigenstate as,
\begin{equation}
E_{n}(\phi)=2t \cos\left[\frac{2\pi}{N}\left(n+\frac{\phi}{\phi_{0}}\right)
\right]
\label{equ11}
\end{equation}
where, $n=0,\pm1,\pm2,\ldots$. In Fig.~\ref{figure2}, we present energy-flux
($E$-$\phi$) characteristics of a perfect ring considering the system size 
$N=10$. It shows that the energy levels vary periodically with $\phi$ 
providing $\phi_0$ flux-quantum periodicity.
At an integer or half-integer flux quantum, each of the energy levels has 
an extremum i.e., either a maximum or a minimum, and therefore, at these 
values of $\phi$ current should disappear. Below, we discuss the 
current-flux ($I$-$\phi$) characteristics for two different cases. In one 
case we take the rings with fixed number of electrons $N_e$, and in the 
other case the rings have some constant chemical potential $\mu$.

\begin{figure}[ht]
{\centering\resizebox*{8cm}{9cm}{\includegraphics{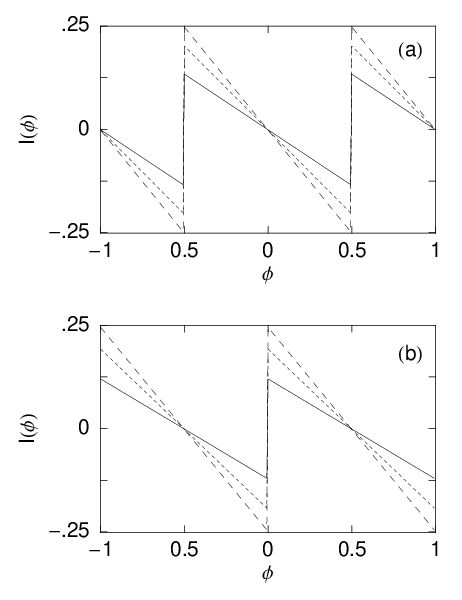}}\par}
\caption{Current-flux characteristics for an ordered one-channel mesoscopic
ring when $N$ is fixed at $50$. The solid, dotted and dashed curves in (a) 
correspond to $N_e=9$, $15$ and $23$, respectively, while in (b) they
represent $N_e=8$, $14$ and $22$, respectively.}
\label{figure3}
\end{figure}

\vskip 0.2cm
\noindent
{\underline{Persistent current: rings with fixed $N_e$}:} The current 
carried by the $n$th energy eigenstate, whose energy is given by 
Eq.~\ref{equ11}, can be obtained through the relation,
\begin{equation}
I_{n}(\phi)=\left(\frac{4\pi t}{N\phi_{0}}\right)\sin\left[\frac{2\pi}{N}
\left(n+\frac{\phi}{\phi_{0}}\right)\right].
\label{equ12}
\end{equation} 
At absolute zero temperature ($T=0$K), total persistent current is 
determined by taking the sum of individual contributions from the 
lowest $N_e$ energy levels. The variation of persistent current as a 
function of flux $\phi$ is presented in Fig.~\ref{figure3}, where (a) 
corresponds to the ring with odd $N_e$, while the results for even 
$N_e$ are shown in (b). From the spectra we observe that the current 
exhibits a saw-tooth like behavior with sharp transitions at half-integer 
and integer flux quanta for the rings having odd and even $N_e$, 
respectively. For all these cases, current varies periodically with 
$\phi$ providing $\phi_0$ periodicity. 

\begin{figure}[ht]
{\centering\resizebox*{8cm}{9cm}{\includegraphics{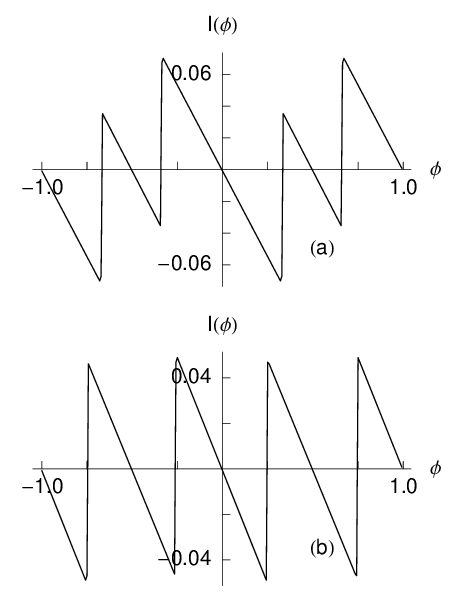}}\par}
\caption{Persistent current as a function of $\phi$ for a ordered mesoscopic
ring, where (a) $\mu=-1$ and (b) $\mu=-1.25$. Here we fix $N=100$.}
\label{figure4}
\end{figure}

\vskip 0.2cm
\noindent
{\underline{Persistent current: rings with fixed $\mu$}:}
When the rings are characterized with fixed chemical potential $\mu$, 
instead of $N_e$, total current at $T=0$K will be obtained by adding 
all individual contributions from the energy levels having energies less 
than or equal to $\mu$. As the chemical potential is fixed, total number 
of electrons varies as a function of AB flux $\phi$ except for some 
typical choices of $\mu$, where the rings contain fixed number of 
electrons. In Fig.~\ref{figure4}, we illustrate current-flux
characteristics for a single-channel mesoscopic ring, described with
constant $\mu$, when the ring is free from impurities. Here we set
$N=100$. Several kink-like structure appears at different values of 
flux $\phi$, associated with the choice of $\mu$ which is clearly 
observed from the $I$-$\phi$ spectra (Fig.~\ref{figure4}). All these
currents oscillate with $\phi$ exhibiting $\phi_0$ periodicity.

\subsubsection{Rings with impurity}

Metals are intrinsically disordered which tends to decrease persistent 
current due to the localization effect~\cite{tvr} of energy eigenstates. 
To address the role of impurities on persistent current now we analyze 
the behavior of current-flux characteristics together with energy band 
structure of some typical mesoscopic rings in presence of disorder.

\begin{figure}[ht]
{\centering\resizebox*{7.75cm}{4.5cm}{\includegraphics{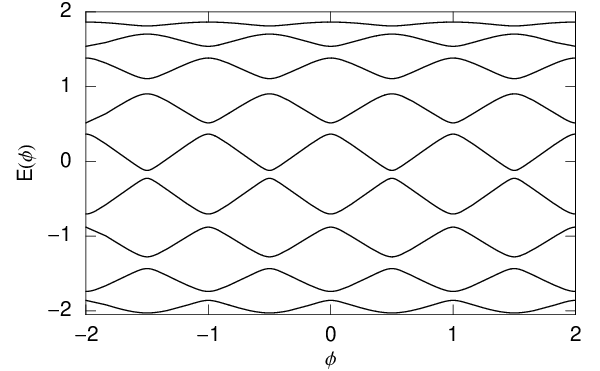}}\par}
\caption{Electron energy levels as a function of the flux $\phi$ for
a one-channel ring ($N=10$) in the presence of impurity with strength 
$W=1$.}
\label{figure5}
\end{figure}

\vskip 0.2cm
\noindent
{\underline{Energy spectra}:} In presence of impurity in a ring, finite 
gaps open between the energy levels at the points of intersection, like
as energy gaps generated in the band structure problem. All these energy
levels vary continuously with flux $\phi$ and no crossing between the  
neighboring levels takes place through the energy band window. In
Fig.~\ref{figure5} we present the energy band structure of a single-channel
ring in presence of diagonal disorder setting $N=10$. The impurities in the 
ring are introduced by selecting the on-site energies, $\epsilon_i$, randomly 
from a ``Box" distribution function of width $W=1$. In presence of impurity
all the degeneracies get removed and energy levels oscillates continuously,
without exhibiting any crossing, result a smooth variation of persistent
current which can be followed from the forthcoming sub-sections.

\begin{figure}[ht]
{\centering\resizebox*{8cm}{9cm}{\includegraphics{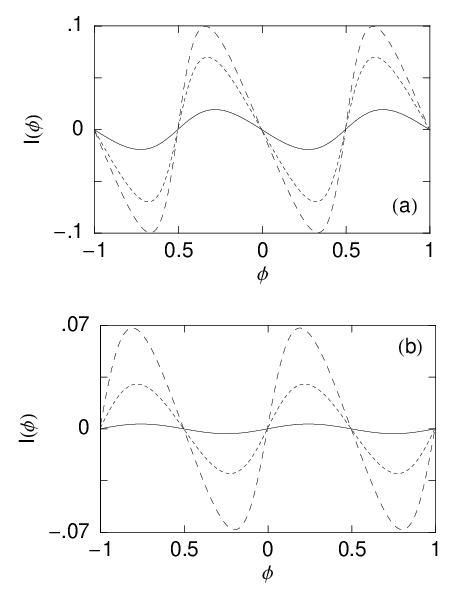}}\par}
\caption{Persistent current as a function of flux $\phi$ for a disordered
ring described with constant $N_e$. The solid, dotted and dashed line in (a)
correspond to $N_e=9$, $15$ and $23$, respectively, while in (b) they 
represent $N_e=8$, $14$ and $22$, respectively. Here we set $N=50$ and 
$W=1$.}
\label{figure6}
\end{figure}

\vskip 0.2cm
\noindent
{\underline{Persistent current: rings with constant $N_e$}:}
The characteristic behavior of persistent current in presence of disorder
is illustrated in Fig.~\ref{figure6} where we set the disorder strength
$W=1$. The results are shown for a $50$-site single-channel ring, where (a)
corresponds to the results for odd $N_e$ and the results for even $N_e$
are given in (b). 
From the spectra it is observed that the current varies continuously 
as a function of $\phi$ with much reduced amplitude compared to the 
impurity-free case (see Fig.~\ref{figure3}). The smooth variation of 
persistent current is clearly followed from the energy band spectrum
since energy levels become continuous as long as the impurities are 
introduced in the ring. On the other hand, the suppression of current 
amplitudes is solely due to the localization effect of the energy 
eigenstates in the presence of impurity. Here all these results are 
described for some typical disordered configurations, and in fact, we 
examine that the qualitative behavior of the persistent currents does not 
depend on the specific realization of disordered configurations. This is 
a generic feature of persistent current for any one-channel 
non-interacting rings in presence of impurity those are described with 
constant $N_e$.

\begin{figure}[ht]
{\centering\resizebox*{8cm}{8cm}{\includegraphics{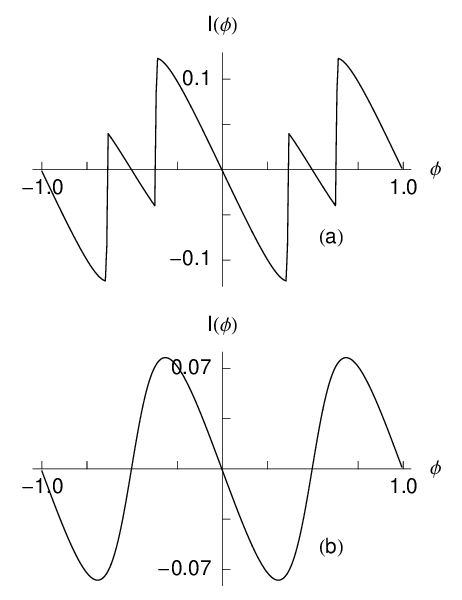}}\par}
\caption{Persistent current as a function of flux $\phi$ for a disordered
ring, where (a) $\mu=-1$ and (b) $\mu=-1.25$. The ring size and impurity
strength are the same as in Fig.~\ref{figure6}.}
\label{figure7}
\end{figure}

\vskip 0.2cm
\noindent
{\underline{Persistent current: rings with constant $\mu$}:}
The situation is somewhat interesting when we fix $\mu$ instead of $N_e$ 
to characterize persistent current in a disordered mesoscopic ring.
As representative example, in Fig.~\ref{figure7} we present the results for
a disordered ring, where we fix $\mu=-1$ in (a) and in (b) $\mu$ is set 
equal to $-1.5$. Interestingly, it is observed that depending on the choice
of $\mu$ persistent current can exhibits a continuous-like or a kink-like
variation with flux $\phi$. In all these cases current oscillates periodically
with $\phi$ showing $\phi_0$ flux-quantum periodicity.

\subsection{Non-interacting multi-channel rings}

In this sub-section we focus our attention on magneto-transport of 
non-interacting electrons in multi-channel mesoscopic rings~\cite{sanu1} 
studied within a simple one-band tight-binding Hamiltonian. A schematic 
\begin{figure}[ht]
{\centering\resizebox*{5cm}{3cm}{\includegraphics{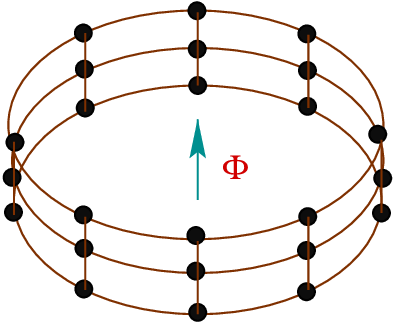}}\par}
\caption{(Color online). Schematic view of a multi-channel cylinder 
threaded by a magnetic flux $\phi$. Filled circles correspond to the 
atomic sites in the cylinder.}
\label{figure8}
\end{figure}
view of such a multi-channel ring geometry threaded by an AB flux $\phi$ 
is shown in Fig.~\ref{figure8}. 
Considering the lattice spacing both in the longitudinal and transverse
directions are identical i.e., the surface of the cylinder forms a square
lattice, we can write the Hamiltonian of the system by the TB formulation 
as,
\begin{equation}
H = \sum_x \epsilon_x c_x^{\dagger}c_x + \sum_{<xx^{\prime}>}\left[
t_{x x^{\prime}} e^{i\theta_{x x^{\prime}}} c_x^{\dagger} c_{x^{\prime}} 
+ t_{x x^{\prime}} e^{-i\theta_{x x^{\prime}}} c_{x^{\prime}}^{\dagger} c_x 
\right] 
\label{equ21-1}
\end{equation}
where, $\epsilon_x$ is the on-site energy of the lattice point $x$ of 
coordinate, say, ($i,j$). $t_{xx^{\prime}}$ is the hopping strength between 
the lattice points $x$ and $x^{\prime}$ and $\theta_{xx^{\prime}}$ is the 
phase factor acquired by an electron due to the longitudinal hopping in 
presence of AB flux $\phi$. The study of persistent currents in such 
multi-channel systems becomes much more relevant compared to strictly 
one-dimensional rings (see Fig.~\ref{figure1}), where we get only one 
channel that carries current, since most of the conventional experiments 
are performed in rings with finite width. Here, we will describe the 
characteristic properties of persistent current together with energy band
spectrum for some non-interacting multi-channel rings concerning the 
dependence of the current on total number of electrons $N_e$, chemical 
potential $\mu$, strength of disorder $W$ and number of channels. 

\subsubsection{Energy band structure}

To elucidate the nature of persistent current in mesoscopic multi-channel 
systems, let us first describe the energy-flux characteristics of a small 
\begin{figure}[ht]
{\centering\resizebox*{8cm}{9cm}{\includegraphics{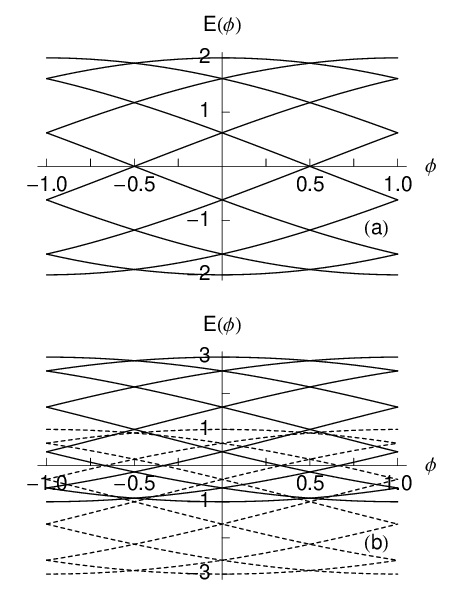}}\par}
\caption{Energy levels as a function of magnetic flux $\phi$, for (a) a 
perfect ring (one layer) with $N=10$ and (b) a perfect cylinder having two 
identical rings taking $N=10$ in each of these two rings.}
\label{figure9}
\end{figure}
cylindrical system that threads a magnetic flux $\phi$. In order to have 
a deeper insight to the problem we begin our discussion with the simplest 
possible system which can be calculated analytically. This is the case of 
a two-layer impurity free cylinder threaded by a magnetic flux $\phi$. This 
cylindrical system can be treated as two one-channel rings placed one above 
the other and they are connected by some vertical bonds (like as in 
Fig.~\ref{figure8}). For a strictly one-dimensional ring i.e., 
for one layer the energy of $n$th eigenstate is expressed in the form
$E_n(\phi) = 2 t\cos\left[\frac{2\pi}{N}\left(n+\frac{\phi}{\phi_0}\right)
\right]$ (see Eq.~\ref{equ11}), where $n=0$, $\pm 1$, $\pm 2$, $\ldots$.
Here $t$ is the nearest-neighbor hopping strength and $N$ is the total 
number of lattice points in the ring. The behavior of such energy levels 
is shown in Fig.~\ref{figure9}(a), where we choose $N=10$. This spectrum
indicates that the energy levels are bounded within the range $-2$ to $2$ 
in the scale of $t$ and the crossing of the energy levels (flux points where 
energy levels have degeneracy) occurs at half-integer or integer multiples 
of $\phi_0$. Now as we add another ring with the previous one-dimensional 
ring and connect it by $N$ vertical bonds it becomes a regular cylinder 
with two identical layers. For such a system, two different energy bands 
of discrete energy levels are obtained, each of which contains $N$ number 
of energy levels and they are respectively expressed in the form,
$E_{1n}(\phi)=-t+2t\cos\left[\frac{2\pi}{N}\left(n+\frac{\phi}{\phi_0}
\right)\right]$ and
$E_{2n}(\phi)=-t+2t\cos\left[\frac{2\pi}{N}\left(n+\frac{\phi}{\phi_0}
\right)\right]$, where the symbol $n$ corresponds to the same meaning as
above and $t$ is the nearest-neighbor hopping strength which is identical 
both for the longitudinal and transverse directions. In Fig.~\ref{figure9}(b),
we show the energy levels of a small impurity-free cylinder considering 
$N=10$, where the solid and dotted line represent the energy levels in 
two separate energy bands, respectively. These two different energy bands 
are bounded respectively in 
the range $-1$ to $3$ and $-3$ to $1$ in the scale of $t$. Accordingly, an
overlap energy region appears for these two energy bands within the range 
$-1$ to $1$, as shown in Fig.~\ref{figure9}(b). In this overlap region, 
energy levels cross each other at several other flux points in addition
to the half-integer and integer multiples of $\phi_0$ which result different
characteristic features of persistent current. For cylinders with more 
than two layers, we get more energy bands like above and therefore several
other overlap energy regions appear in energy band spectrum. 

In presence of impurity in multi-channel cylinders, gaps open at crossing 
points of energy levels and they provide a continuous-like variation with
AB flux $\phi$, similar to the case of conventional one-dimensional rings
with impurities (Fig.~\ref{figure5}). In all these perfect and disordered 
multi-channel cylinders, energy levels vary periodically with period $\phi_0$.

Below, we will investigate the behavior of persistent current in small 
cylindrical systems and we believe that the results might be quite helpful 
to explain characteristic properties of persistent current in larger system 
sizes, even in more complex geometries.

\subsubsection{Persistent current}

\begin{figure}[ht]
{\centering\resizebox*{8cm}{10cm}{\includegraphics{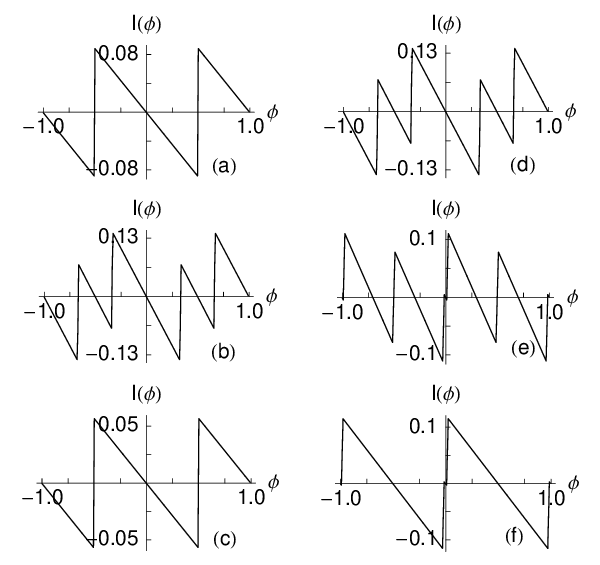}}\par}
\caption{Persistent current as a function of AB flux $\phi$ for an ordered
mesoscopic cylinder with two layers in which each layer contains $100$ atomic
sites. The first column corresponds to the results for constant $N_e$, where 
(a) $N_e=25$, (b) $N_e=100$ and (c) $N_e=185$, while the second column 
describes the results for fixed $\mu$, where (d) $\mu=0$, (e) $\mu=-0.5$ 
and (f) $\mu=-1.5$.}
\label{figure10}
\end{figure}

\noindent
{\underline{Perfect cylinders}:}
Here we describe the phenomenon of persistent current for some impurity-free 
multi-channel systems of cylindrical geometry containing  two layers concerning 
the dependence of current on total number of electrons $N_e$ and chemical 
potential $\mu$. As illustrative example, in Fig.~\ref{figure10} we show 
the variation of persistent current as a function of AB flux $\phi$ for
an ordered cylinder taking $N=100$ in each of these two layers. The first 
column corresponds to the case where the cylinder is characterized with
constant $N_e$, while the second column represents the results where we
fix $\mu$ instead of $N_e$. To reveal the effect of energy overlap region 
on persistent current here we present our results for three different values 
of $N_e$. Figures~\ref{figure10}(a) and (c) correspond to the currents 
when $N_e=25$ (low) and $N_e=185$ (high), respectively, while the current
for the intermediate value of $N_e$ ($N_e=100$) is shown in 
Fig.~\ref{figure10}(b). Both for the low and high values of $N_e$, we notice
that the current gets saw-tooth like nature with sharp transitions only at 
half-integer multiples of $\phi_0$, similar to that of conventional 
one-channel impurity-free rings with odd $N_e$ (Fig.~\ref{figure3}(a)).  
This feature can be explained as follows. When $N_e$ is set at $25$,
the net current is available by taking the contributions from lowest $25$
energy levels those lie well below the energy overlap region. Now, away from 
this overlap region, the energy levels behave exactly the same way with 
perfectly single-channel rings, which results a similar kind of saw-tooth 
shape as obtained in one-channel rings. On the other hand, when the filling
factor $N_e$ is fixed at $185$ the situation is somewhat different than 
the earlier one. To obtain net current for this case we cross the overlap 
energy region since the highest energy level that contributes current lies far 
above this overlap region and the net contribution to the current from the 
energy levels within this overlap region vanishes and therefore no new 
feature appears in persistent current compared to the system with $N_e=25$. 
Finally, we see that for the intermediate value of $N_e$ ($N_e=100$) 
persistent current gives some additional kink-like structures across 
$\phi=\pm 0.5$. These kinks 
\begin{figure}[ht]
{\centering\resizebox*{8cm}{7cm}{\includegraphics{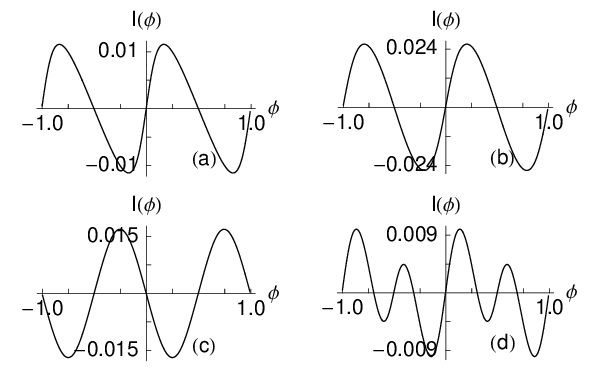}}\par}
\caption{Current-flux characteristics for the disordered ($W=1$) cylinders 
with two layers taking $N=100$ in each layer. The results for the fixed 
$N_e$ are plotted in the first column, where (a) $N_e=50$ and (b) $N_e=100$, 
while the results for the constant $\mu$ are plotted in the second column, 
where (c) $\mu=0$ and (d) $\mu=-0.5$.}
\label{figure11}
\end{figure}
are due to the different contributions of energy levels in the overlap 
energy region, since in this region energy levels have more degeneracies 
at different flux points in addition to the half-integer and integer 
multiples of $\phi_0$. For other multi-channel systems with more than two 
layers, additional kinks may appear in persistent current at different 
values of $\phi$ depending on the choice of $N_e$ and number of layers.

In this cylindrical system where we fix chemical potential $\mu$ instead 
$N_e$, different kink-like structures in persistent current are also
appeared as visible from our results shown in the second column of 
Fig.~\ref{figure10}. All these cases current exhibits $\phi_0$ flux-quantum 
periodicity.

\vskip 0.2cm
\noindent
{\underline{Disordered cylinders}:}
At the end of this sub-section, we describe the effect of impurities on
persistent current in a multi-channel system. The results shown
in Fig.~\ref{figure11} and they are computed for the same system size as
taken above. The impurities are introduced by selecting the site energies 
($\epsilon_x$) randomly from a ``Box" distribution function of width 
$W=1$. All these numerical results are performed for different isolated 
disordered configurations. The results for constant $N_e$ are shown in the
first column, while in the second column the results for fixed $\mu$ are
presented. In all these cases persistent current varies continuously, 
without providing any kink-like structure, as a function of flux $\phi$ 
exhibiting $\phi_0$ flux-quantum periodicity. Another significant point 
observed from Figs.~\ref{figure11}(a) and (b) is that the slope of the 
current in zero-field limit ($\phi \rightarrow 0$) changes in opposite 
direction though in these two cases the cylinder contains even number 
of electrons. This emphasizes the signature of random sign of low-field 
currents in multi-channel systems. The detailed description of low-field 
magnetic response will be available in Section $V$. 

Thus we can summarize that the behavior of persistent current in 
multi-channel systems strongly depends on total number of electrons 
$N_e$, chemical potential $\mu$, total number of channels and disordered 
configurations.

\section{Effect of electron-electron correlation on persistent current
in one-channel rings}

The phenomenon of persistent current in small conducting ring threaded
by a magnetic flux $\phi$ was predicted in the pioneering work of
B\"{u}ttiker, Imry and Landauer~\cite{butt}, and since then it has been
discussed in several theoretical papers~\cite{butt1,cheu1,cheu2,mont,mont1,
land,byers,von,bouc,alts,schm,abra,mull}. Although this phenomenon is 
thought to be qualitatively understood within the framework of one-electron 
picture~\cite{cheu1,cheu2,mont,mont1,land,byers,von,bouc,alts,schm}, but it 
fails to explain many experimental results~\cite{levy,deb,reu,jari,chand,
mailly1}. 
A typical example of such discrepancy between theory and experiment 
is that the amplitudes of measured persistent currents are orders of 
magnitude larger than theoretical predictions. It is generally believed 
that electron-electron correlation and disorder have major role on the 
enhancement of persistent current, but no consensus has yet been reached. 
Another important controversial issue is that, experimentally both $\phi_0$ 
and $\phi_0/2$ periodicities have been observed and it is also found that the 
$\phi_0/2$ oscillations near zero magnetic flux exhibit diamagnetic response. 
The explanation of these results in terms of the ensemble averaged 
persistent currents is also quite intriguing, and the calculations show that 
the disordered averaged current crucially depends on the choice of the 
ensemble~\cite{cheu1,mont}. The explanation of experimental results become 
even much more illusive since recently Kravtsov {\em et al.}~\cite{krav}
have shown that additional currents may be generated in rings by other 
mechanisms which are not experimentally distinguishable from the persistent 
current.

\subsection{Impurity free rings}

To emphasize the precise role of electron-electron correlation, in this 
sub-section we focus on the exact calculation of persistent current and 
Drude weight in perfect one-dimensional Hubbard rings~\cite{sanu2} threaded 
by a magnetic flux $\phi$. Our study reveals that, certain aspects of the 
many-body effects on persistent current have not been investigated in the 
literature clearly, and here we will show that the Hubbard correlation
leads to many significant effects. We restrict ourselves to small Hubbard 
rings and these results might also be helpful to understand the physical 
properties of TTF-TCMQ conductors, various aromatic molecules and systems 
of connected quantum dots\cite{yu}. With the new advancements of the 
nanoscience and technology, it is now possible quite simply to fabricate 
such small rings, and in a recent experiment Keyser {\em et al.}~\cite{keyser} 
reported the evidence of anomalous Aharonov-Bohm oscillations from the 
transport measurements on small quantum rings with less than ten electrons. 
The electron-electron interaction becomes much more important in these small 
rings with very few number of electrons since the Coulomb potential is not 
screened much, and we will show that the electron-electron interaction 
provides similar anomalous oscillations in persistent current as a function 
of magnetic flux $\phi$. 

We use the Hubbard model to represent the system which for a $N$-site ring 
enclosing a magnetic flux $\phi$ can be written in this form,
\begin{eqnarray}
H & = & t\sum_{\sigma}\sum_{i=1}^N \left[e^{i \theta} c_{i,\sigma}^{\dagger} 
c_{i+1,\sigma}+ e^{-i\theta} c_{i+1,\sigma}^{\dagger} c_{i,\sigma} \right] 
\nonumber \\
 & + & U \sum_{i=1}^N n_{i\uparrow}n_{i\downarrow}
\label{equ13}
\end{eqnarray}
where $c_{i \sigma}^{\dagger}$ ($c_{i \sigma}$) is the creation (annihilation)
operator and $n_{i \sigma}$ is the number operator for an electron in the
Wannier state $|i \sigma>$. The parameters $t$ and $U$ are the 
nearest-neighbor hopping integral and the strength of the Hubbard 
correlation, respectively. The phase factor $\theta=2\pi \phi/N$ appears 
due to the flux $\phi$ threaded by the ring. Throughout this sub-section 
we set $t=-1$.

\subsubsection{Persistent current and energy band structure}

At absolute zero temperature ($T=0$K), persistent current in a ring 
threaded by a magnetic flux $\phi$ is obtained through the 
expression~\cite{cheu2},
\begin{equation}
I(\phi)=-\frac{\partial E_0(\phi)}{\partial \phi}
\label{equ14}
\end{equation}
where, $E_0(\phi)$ is the ground state energy. We determine this quantity 
exactly to understand unambiguously the role of electron-electron 
interaction on persistent current, and this is obtained by exact numerical 
diagonalization of the many-body Hamiltonian (Eq.~\ref{equ13}).

\vskip 0.2cm
\noindent
{\underline{Rings with two opposite spin electrons}:}
To have a deeper insight to the problem, we first consider two electron 
systems and begin our study with the simplest possible system which can be 
treated analytically up to certain level. This is the case of a three-site
\begin{table}
\caption{Eigenvalues ($\lambda$) and eigenstates of two opposite spin 
electrons for N=3.}
\label{table}
~\\
\begin{tabular}{|c|c|c|c|}
\hline \hline
Total & \multicolumn{3}{c|}{$U=0$} \\ \cline{2-4}
spin $S$ & $\lambda$ & Degeneracy & Eigenstate \\
\hline
 & $-4$ & $1$ & $\left(-\frac{1}{\sqrt 2}, -\frac{1}{\sqrt 2}, -\frac{1}
{\sqrt 2}, -1, -1, 1 \right)$ \\ \cline{2-4}
 & $-1$ & $2$ & $\left(0, \sqrt 2, -\sqrt 2, 1, 0, 1 \right)$ \\ 
 & & & $\left(-\sqrt 2, 0, \sqrt 2, -1, 1, 0 \right)$ \\ \cline{2-4}
$0$ & $2$ & $3$ & $\left(\frac{1}{\sqrt 2}, 0, \frac{1}{\sqrt 2}, 0, 0, 1 
\right)$ \\ 
 & & & $\left(0, -\frac{1}{\sqrt 2}, - \frac{1}{\sqrt 2}, 0, 1, 0 \right)$ \\ 
 & & & $\left(-\frac{1}{\sqrt 2}, - \frac{1}{\sqrt 2}, 0, 1, 0, 0 \right)$ \\
\cline{1-4} 
 & $-1$ & $2$ & $\left(1, 0, 1 \right)$ \\
 $1$ & & & $\left(-1, 1, 0 \right)$ \\
 & $2$ & $1$ & $\left(-1, -1, 1 \right)$ \\ \cline{1-4}
\hline
\end{tabular}
\vskip 0.1cm
\begin{tabular}{|c|c|c|c|}
\hline
Total & \multicolumn{3}{c|}{$U=2$} \\ \cline{2-4}
spin $S$ & $\lambda$ & Degeneracy & Eigenstate \\
\hline
 & $-2 \sqrt 3$ & $1$ & $~~~~\left(a, a, a, -1, -1, 1 \right)~~~~$ \\ 
\cline{2-4}
 & $0$ & $2$ & $\left(0, \frac{1}{\sqrt 2}, -\frac{1}{\sqrt 2}, 1, 0, 1 
\right)$ \\ 
 & & & $\left(-\frac{1}{\sqrt 2}, 0, \frac{1}{\sqrt 2}, -1, 1, 0 \right)$ 
\\ \cline{2-4}
$0$ & $3$ & $2$ & $\left(0, -\sqrt 2, \sqrt 2, 1, 0, 1 \right)$ \\ 
 & & & $~~\left(\sqrt 2, 0, -\sqrt 2, -1, 1, 0 \right)~$ \\ 
 & $2 \sqrt 3$ & $1$ & $\left(b, b, b, -1, -1, 1 \right)$ \\
\cline{1-4} 
 & $-1$ & $2$ & $\left(1, 0, 1 \right)$ \\
 $1$ & & & $\left(-1, 1, 0 \right)$ \\
 & $2$ & $1$ & $\left(-1, -1, 1 \right)$ \\ \cline{1-4}
\hline \hline
\end{tabular}
\end{table}
ring with two opposite spin ($\uparrow,\downarrow$) electrons. The
total Hamiltonian of this system becomes a ($9 \times 9$) matrix which 
can be block diagonalized to two sub-matrices by proper choice of the 
basis states. The orders of the two sub-matrices are ($6 \times 6$) and
($3 \times 3$), respectively. This can be achieved by constructing the basis 
states for each sub-space with a particular value of the total spin $S$.
The basis set ($\mathcal{A}$) for the six-dimensional sub-space, spanned for 
$S=0$, is chosen as:
\[\mathcal{A} \equiv \left\{\begin{array}{ll}
  |\uparrow\downarrow,0,0 > \\
  |0,\uparrow\downarrow,0 > \\
  |0,0,\uparrow\downarrow > \\
  \frac{1}{\sqrt 2} \left(|\uparrow,\downarrow,0> - |\downarrow,\uparrow,0> 
  \right) \\
  \frac{1}{\sqrt 2} \left(|0,\uparrow,\downarrow> - |0,\downarrow,\uparrow> 
  \right) \\
  \frac{1}{\sqrt 2} \left(|\downarrow,0,\uparrow> - |\uparrow,0,\downarrow> 
  \right) 
     \end{array}
  \right\}~~S=0 \]
On the other hand, the other basis set ($\mathcal{B}$) for the 
three-dimensional sub-space, spanned for $S=1$, is chosen as:
\[\mathcal{B} \equiv \left\{\begin{array}{ll}
  \frac{1}{\sqrt 2} \left(|\uparrow,\downarrow,0> + |\downarrow,\uparrow,0> 
  \right) \\
  \frac{1}{\sqrt 2} \left(|0,\uparrow,\downarrow> + |0,\downarrow,\uparrow> 
  \right) \\
  \frac{1}{\sqrt 2} \left(|\uparrow,0,\downarrow> + |\downarrow,0,\uparrow> 
  \right) 
     \end{array}
  \right\}~~S=1 \]
In absence of any magnetic field, we list the energy eigenvalues, 
eigenstates, and the degeneracy of the levels of this system in 
Table~\ref{table} for $U=0$ and $U=2$.
In this table we set $a=-\sqrt 2/\left(1+\sqrt 3\right)$ and
$b=\sqrt 2/\left(-1+\sqrt 3\right)$. Both for $U=0$ and $U \ne 0$ cases, the 
two-fold degeneracy gets lifted in the presence of magnetic flux $\phi$.
It is apparent from this table that the eigenvalues, eigenstates and also
the degeneracy of the energy levels are not affected by the correlation in
the three-dimensional sub-space. This is due to the fact that the basis
set $\mathcal {B}$ does not involve any doubly occupied state but this 
is not true in the other sub-space. The insertion of the magnetic flux
does not alter the above structure of the Hamiltonian though it becomes
now field dependent.

\begin{figure}[ht]
{\centering\resizebox*{8cm}{10cm}{\includegraphics{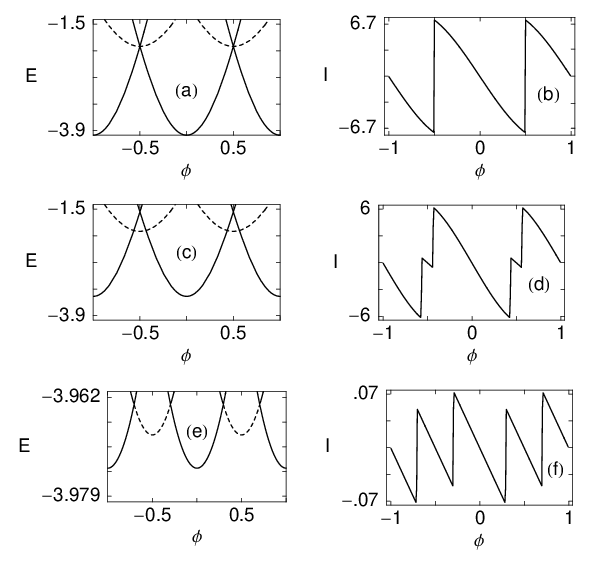}}\par}
\caption{$E$-$\phi$ and $I$-$\phi$ characteristics of the following two 
opposite spin ($\uparrow,\downarrow$) electron systems: (i) $N=3$, $U=0$ in 
(a) $\&$ (b); (ii) $N=3$, $U=2$ in (c) $\&$ (d); and (iii) $N=25$, $U=3$ in 
(e) $\&$ (f).}
\label{figure12}
\end{figure}

In Fig.~\ref{figure12}(a), we show some of the low-lying energy levels 
$E(\phi)$'s as a function of magnetic flux $\phi$ of this three-site ring 
in absence of any electron correlation ($U=0$), and the corresponding 
$I(\phi)$ versus $\phi$ curve is plotted in Fig.~\ref{figure12}(b).
The $E$-$\phi$ and $I$-$\phi$ curves for $U=2$ case are given in 
Figs.~\ref{figure12}(c) and (d), respectively. The dotted curves in 
Figs.~\ref{figure12}(a) and (c) correspond to the $U$-independent energy 
levels as mentioned above. Quite interestingly, from Fig.~\ref{figure12}(c) 
we see that even in presence of interaction, one of the $U$-independent 
energy levels becomes the ground state energy level in certain intervals 
of $\phi$ (e.g., around $\phi=\pm 0.5$). In other regions of $\phi$, the 
ground state energy increases due to the correlation and the $E_0$-$\phi$ 
curves become much shallow, as expected. As a result, the usual saw-tooth 
shape of current-flux characteristics for $U=0$ changes drastically as 
plotted in Fig.~\ref{figure12}(d) and the role of e-e correlation is quite 
complex rather than a simple suppression of persistent current as predicted 
earlier. A sudden change in the direction and magnitude of persistent current 
occurs, solely due to the correlation, around $\phi=\pm 0.5$ and it forms 
kink-like structures in the current, as illustrated in Fig.~\ref{figure12}(d). 
The kinks get wider with increasing the strength of $U$, and the most 
significant result is that the current remains invariant inside the kinks 
irrespective of the correlation strength. It is notice seen that, the 
correlation does not affect $\phi_{0}$ flux periodicity of persistent 
current.

From this above background we can investigate the role of electronic
correlation on persistent current in mesoscopic rings with larger sizes. 
\begin{figure}[ht]
{\centering\resizebox*{8cm}{10cm}{\includegraphics{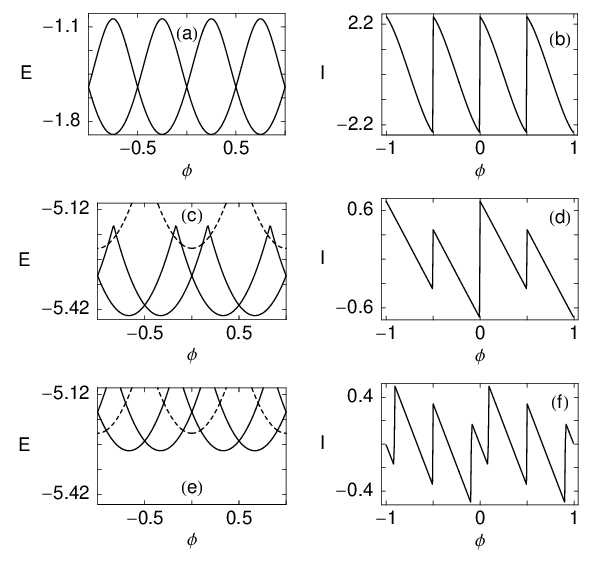}}\par}
\caption{$E$-$\phi$ and $I$-$\phi$ characteristics of the following 
three ($\uparrow,\uparrow,\downarrow$) electron systems: (i) $N=3$, $U=3$ in 
(a) $\&$ (b); (ii) $N=10$, $U=3$ in (c) $\&$ (d); and (iii) $N=10$, $U=15$ 
in (e) $\&$ (f).}
\label{figure13}
\end{figure}
In Figs.~\ref{figure12}(e) and (f), we display $E$-$\phi$ and $I$-$\phi$ 
characteristics for the ring containing two opposite spin electrons with 
$N=25$ and $U=2$, respectively. These figures resemble Figs.~\ref{figure12}(c) 
and (d), respectively, and it implies that correlation plays the same role 
in both the cases. Therefore, we can precisely conclude that away from 
half-filling, a ring consisting of two opposite spin electrons always 
exhibits kinks in persistent current for any non-zero value of $U$.

\vskip 0.2cm
\noindent
{\underline{Rings with two up and one down spin electrons}:}
Now we consider rings with two up and one down spin electrons as 
illustrative examples of three spin electron systems. We plot the $E$-$\phi$ 
and $I$-$\phi$ curves for the half-filled band case ($N=3$ and $n=3$, where 
$n$ denotes the number of electrons) with $U=3$ in Figs.~\ref{figure13}(a)
and (b), respectively. We find that the correlation just diminishes the 
magnitude of the current compared to that of the noninteracting case. In 
Figs.~\ref{figure13}(c) and (d), we respectively plot the $E$-$\phi$ and 
$I$-$\phi$ curves of a non-half-filled ring with $N=10$ and $U=3$, while 
Figs.~\ref{figure13}(e) and (f) are these curves for the $U=15$ case. The 
behavior of persistent current as a function of flux $\phi$ are quite 
different at low and high values of $U$. From Fig.~\ref{figure13}(c), it 
is evident that for the low value of $U$ all $U$-independent energy levels 
\begin{figure}[ht]
{\centering\resizebox*{8cm}{10cm}{\includegraphics{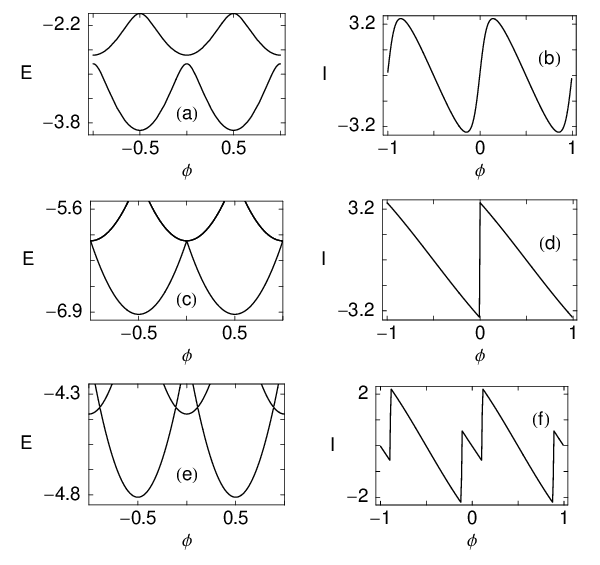}}\par}
\caption{$E$-$\phi$ and $I$-$\phi$ characteristics of the following four 
($\uparrow,\uparrow,\downarrow,\downarrow$) electron systems: (i) $N=4$, $U=2$ 
in (a) $\&$ (b); (ii) $N=6$, $U=0$ in (c) $\&$ (d); and (iii) $N=6$, $U=6$ 
in (e) $\&$ (f).}
\label{figure14}
\end{figure}
(dotted curves) always lie above the ground state energy of the system, and 
Fig.~\ref{figure13}(d) shows that apart from a reduction of persistent 
current the $I$-$\phi$ curve looks very similar to that of a ring without 
any interaction. On the other hand, from Fig.~\ref{figure13}(e) it is 
observed that for the high value of $U$ one of the $U$-independent energy 
levels becomes the ground state energy in certain intervals of $\phi$ 
(e.g., around $\phi=0$) and this produces kinks in the $I$-$\phi$ 
characteristics as depicted in Fig.~\ref{figure13}(f). Certainly there 
exists a critical value $U_c$ of the correlation above which the kinks 
appear in the $I$-$\phi$ characteristics. These features of the energy 
spectra and persistent currents are the 
characteristics of any non-half-filled ring with two up and one down spin 
interacting electrons. Here we note that the half-filled rings exhibit 
$\phi_{0}/2$ periodic currents, while the non-half-filled rings have 
$\phi_{0}$ periodicity.

\vskip 0.2cm
\noindent
{\underline{Rings with two up and two down spin electrons}:}
Next we consider rings with two up and two down spin electrons as 
representative examples of four electron system. Let us first describe the 
half-filled band case i.e., $N=4$ and $n=4$. In absence of any electron 
correlation, sharp discontinuity appears in persistent current at certain 
values of $\phi$. However, the effect of the correlation is quite dramatic 
and it makes the current a continuous function of the flux as given in 
Fig.~\ref{figure14}(b). This is a quite fascinating result since the 
\begin{figure}[ht]
{\centering\resizebox*{8cm}{10cm}{\includegraphics{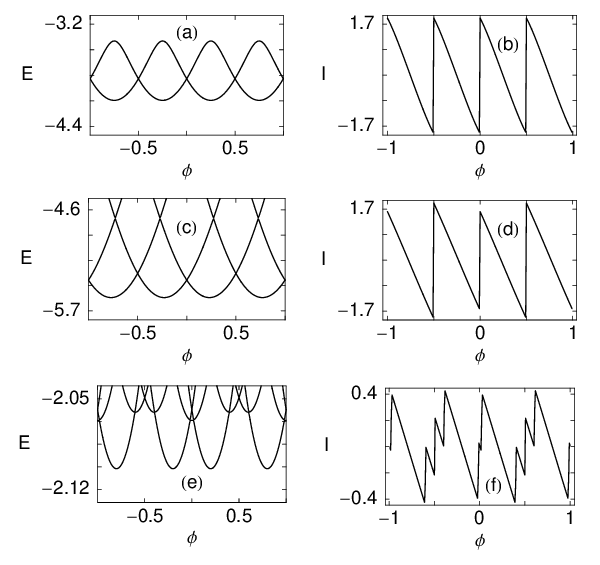}}\par}
\caption{$E$-$\phi$ and $I$-$\phi$ characteristics of the following five 
($\uparrow,\uparrow,\uparrow,\downarrow,\downarrow$) electron systems: 
(i) $N=5$, $U=2$ in (a) $\&$ (b); (ii) $N=6$, $U=2$ in (c) $\&$ (d); and 
(iii) $N=6$, $U=120$ in (e) $\&$ (f).}
\label{figure15}
\end{figure}
correlation drastically changes the analytic behavior of $I(\phi)$, and 
here we will observe that this result also holds true for the other 
half-filled systems with even number of electrons. Away from the 
half-filling we study an interesting typical case of a six-site ring with 
two up and two down spin electrons. This can be considered as a doubly 
ionized benzene-like ring, a system with the promise of experimental 
verification of our predictions. Figures~\ref{figure14}(c) and (d) show 
$E$-$\phi$ and $I$-$\phi$ characteristics with $U=0$, respectively, while 
Figs.~\ref{figure14}(e) and (f) are those plots with $U=6$. It is observed 
from Fig.~\ref{figure14}(f) that the kinks appear in the $I$-$\phi$ curve 
(e.g., around $\phi=0$) for any non-zero value of $U$. Here we interestingly 
note that the kinks are now due to the $U$-dependent eigenstates. Both the 
half-filled and non-half-filled rings exhibit $\phi_0$ periodic currents.

\vskip 0.2cm
\noindent
{\underline{Rings with three up and two down spin electrons}:}
For specific cases of five electron systems we investigate rings with 
three up and two down spin electrons. In Figs.~\ref{figure15}(a) and (b), 
we plot the $E$-$\phi$ and $I$-$\phi$ curves, respectively for the 
non-half-filled ring ($N=5$ and $n=5$). The magnitude of the current
is just diminished due to e-e electron correlation and we find $\phi_0/2$
periodicity. As a non-half-filled five electron case, here we take a singly
ionized benzene-like ring ($N=6$ and $n=5$) and determine the persistent 
\begin{figure}[ht]
{\centering\resizebox*{8cm}{6cm}{\includegraphics{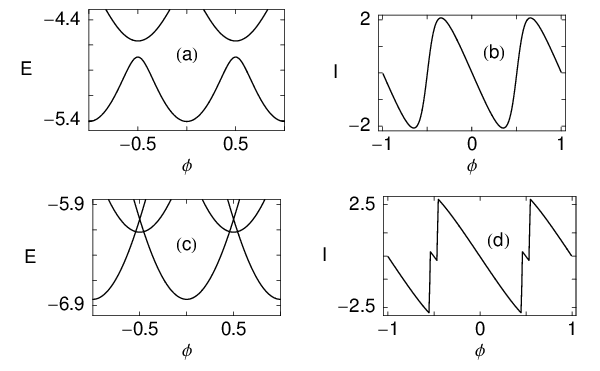}}\par}
\caption{$E$-$\phi$ and $I$-$\phi$ characteristics of the following six 
($\uparrow,\uparrow,\uparrow,\downarrow,\downarrow,\downarrow$) electron 
systems: (i) $N=6$, $U=2$ in (a) $\&$ (b) and (ii) $N=7$, $U=2$ in 
(c) $\&$ (d).}
\label{figure16}
\end{figure}
current both for low and high values of $U$. In Figs.~\ref{figure15}(c) and
(d), we respectively draw the $E$-$\phi$ and $I$-$\phi$ curves considering 
the correlation strength $U=2$, while these diagrams are plotted in 
Figs.~\ref{figure15}(e) and (f) with $U=120$. From Figs.~\ref{figure15}(d) 
and (f), it is clear that like
three electron rings kinks appear in persistent current only after a 
critical value of $U$. This result emphasizes that the characteristic 
features of persistent current in mesoscopic Hubbard rings with odd 
number of electrons are almost invariant. 

\vskip 0.2cm
\noindent
{\underline{Rings with three up and three down spin electrons}:}
Finally, we take six electron systems and study the nature of persistent 
current in rings with three up and three down spin electrons. At half-filling
(a benzene-like ring with $N=6$ and $n=6$), we see that the current becomes
a continuous function of flux $\phi$ as plotted in Fig.~\ref{figure16}(b),
exactly similar to the half-filled four electron case. The $E$-$\phi$ and
$I$-$\phi$ curves for a typical non-half-filled ring with $N=7$ and $n=6$ are
presented in Figs.~\ref{figure16}(c) and (d), respectively.
Here we find striking similarity in the behavior of persistent current 
with other non-half-filled systems containing even number of electrons.
Therefore, it becomes apparent that mesoscopic Hubbard rings with even 
number of electrons exhibit similar characteristic features in 
persistent current.

\subsubsection{Drude weight}

In this sub-section we investigate the response of mesoscopic Hubbard 
rings to a uniform time-dependent electric field in terms of Drude 
weight~\cite{scal1,scal2} $D$, a closely related parameter that 
characterizes the conducting nature of a system as originally predicted 
by Kohn~\cite{kohn}. The Drude weight can be evaluated from the following 
expression~\cite{bouz},
\begin{equation}
D = \frac{N}{4 \pi^2} \left[\frac{\partial^2 E_0(\phi)}{\partial \phi^2} 
\right]_{\phi=\phi_{\mbox{\tiny min}}}
\label{equ15}
\end{equation}
where, $\phi_{\mbox{\tiny min}}$ provides the location of the minimum of 
energy $E_0(\phi)$. A metallic phase is characterized by a finite non-zero 
value of $D$, while it reaches to zero in an insulating 
phase~\cite{kohn}. We show the variation of Drude weight $D$ as a function 
\begin{figure}[ht]
{\centering\resizebox*{8cm}{9cm}{\includegraphics{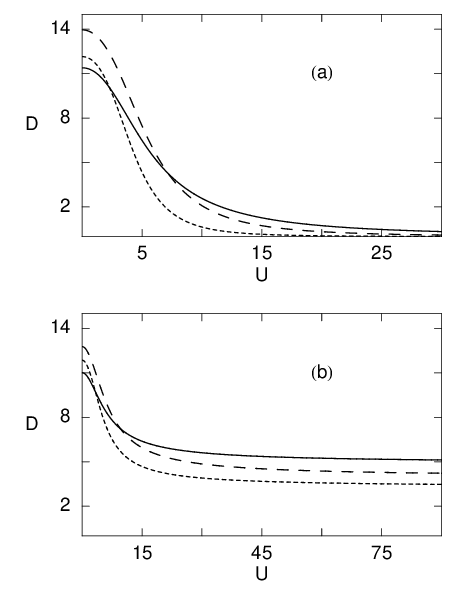}}\par}
\caption{Drude weight ($D$) versus the Hubbard correlation strength ($U$)
for (a) half-filled and (b) non-half-filled band systems. The solid, 
dotted and dashed curves correspond to $3$, $4$ and $5$ electron cases,
respectively.}
\label{figure17}
\end{figure}
of the Hubbard correlation strength $U$ for the half-filled and 
non-half-filled rings with $n=3$, $4$ and $5$ in Figs.~\ref{figure17}(a)
and (b), respectively. For the non-half-filled band cases, the number 
of sites corresponding to a given value of $n$ is taken as $N=n+1$,
and we observe that the other choices of $N$ do not affect the basic
characteristics of the $D$-$U$ curves. For low values of $U$, the 
half-filled systems are in the metallic phase which is clearly evident from 
Fig.~\ref{figure17}(a), and the systems become insulating only when the
correlation strength $U$ becomes very large. In the insulating phase ground 
state does not favor any empty site, and accordingly, the situation is 
somewhat analogous to Mott localization in one-dimensional infinite lattices. 
On the other hand, Fig.~\ref{figure17}(b) shows that the non-half-filled 
rings are always conducting irrespective of the correlation strength $U$.

Throughout our analysis of persistent current in one-channel Hubbard rings 
without any impurity we get several interesting new results. The main 
results are: the appearance of kinks in persistent current, observation of 
both the $\phi_0$ and $\phi_0/2$ periodicities in persistent current, no
singular behavior of persistent current in the half-filled rings with even
number of electrons, evidence of $U$-independent eigenstates, existence of 
both metallic and insulating phases, etc. We also observe discontinuities 
in persistent current at non-integer values of $\phi_0$ due to the electron 
correlation which crucially depends on filling of the ring and also on the 
parity of the number of electrons. This corresponds to the anomalous 
Aharonov-Bohm oscillations in persistent current with much reduced period 
where periodicity is not perfect, and Keyser {\em et al.}~\cite{keyser} 
experimentally observed similar anomalous Aharonov-Bohm oscillations in 
conductance of a few-electron quantum ring.

\subsection{Rings with impurities}

To explore the role of disorder and electron-electron correlation on 
persistent current, in this sub-section, we focus our attention on certain 
systems which closely resemble to the disordered systems where we do not 
require any configuration averaging~\cite{sanu3}. These are chemically 
modulated structures possessing well-defined long-range order and as 
specific examples we consider aperiodic and ordered binary alloy rings. 
We restrict ourselves to small one-dimensional rings, as in the previous 
sub-section, where persistent current can be calculated exactly. We get 
many interesting new results as a consequence of electron-electron 
interaction and disorder. One such promising result is the enhancement of 
persistent current amplitude in these systems due to the electronic 
correlation. This study might also be helpful to understand physical 
properties of benzene-like rings and other aromatic compounds in presence 
of magnetic flux $\phi$.

\subsubsection{Ordered binary alloy rings}

\begin{figure}[ht]
{\centering\resizebox*{5.5cm}{3cm}{\includegraphics{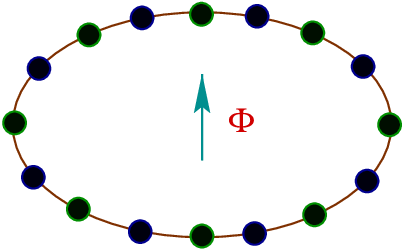}}\par}
\caption{(Color online). Schematic view of a one-dimensional tight-binding
ordered binary alloy ring threaded by a magnetic flux $\phi$. The green and 
blue circles correspond to two different types of atoms.}
\label{figure18}
\end{figure}

Here we investigate current-flux characteristics in an ordered binary
alloy ring (Fig.~\ref{figure18}) at absolute zero temperature ($T=0$K). 
A simple TB Hamiltonian is used to describe the system, and for a $N$-site 
ring it reads,
\begin{eqnarray}
H & = & \sum_{\sigma}\sum_{i=1,3,\ldots}^{N-1} \left(\epsilon_{A} 
n_{i,\sigma} + \epsilon_{B} n_{i+1,\sigma}\right) \nonumber \\
 & + &  
t\sum_{\sigma}\sum_{i=1}^{N} \left(c_{i,\sigma}^{\dagger}c_{i+1,\sigma} 
e^{i\theta} + c_{i+1,\sigma}^{\dagger}c_{i,\sigma}e^{-i\theta} \right) 
\nonumber \\
 & + & U\sum_{i=1}^{N} n_{i\uparrow}n_{i\downarrow}
\label{equ16}
\end{eqnarray}
where, $\epsilon_A$ and $\epsilon_B$ are the on-site potentials for the 
$A$ and $B$ type atoms. All the other symbols of this Hamiltonian carry 
the same meaning as described earlier.
We always choose $N$ to be even to preserve the perfect binary ordering of 
the two types of atoms in the ring, as shown in Fig.~\ref{figure18}. 

Let us now discuss the behavior of persistent current in an ordered 
binary alloy ring and investigate the role of 
\begin{figure}[ht]
{\centering\resizebox*{8cm}{9cm}{\includegraphics{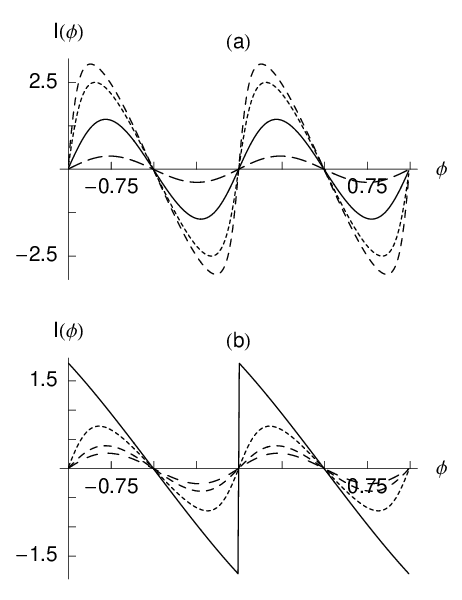}}\par}
\caption{Current-flux characteristics of four ($\uparrow,\uparrow,\downarrow,
\downarrow$) electron ordered binary alloy rings, where (a) $N=4$, $N_e=4$ 
and (b) $N=8$, $N_e=4$. The solid, dotted, small dashed and dashed lines in 
(a) correspond to $U=0$, $2$, $4$ and $10$, respectively, while in (b) they 
are for $U=0$, $2$, $4$ and $6$, respectively.}
\label{figure19}
\end{figure}
electron-electron correlation on persistent current. In a pure ring 
consisting of either $A$ or $B$ type of atoms and in absence of any 
electron correlation, the current shows discontinuity at certain points 
of $\phi$ due to ground state degeneracy and the $I$-$\phi$ curve gets 
a saw-tooth like behavior~\cite{cheu2}. This discontinuity completely 
disappears in a binary alloy ring with $U=0$, as illustrated by the solid 
line in Fig.~\ref{figure19}(a). This is due to the fact that, the binary alloy 
configuration may be considered as a perturbation over the pure ring which 
lifts the ground state degeneracy, and accordingly, the current $I(\phi)$ 
becomes a continuous function of magnetic flux. As we consider 
electron-electron interaction, current always decreases with the 
increase of interaction strength. However, depending on the number 
of electrons $N_e$ in the ordered binary alloy rings, we get enhancement 
of persistent current for the low values of $U$, but it decreases
eventually when $U$ becomes very large. This type of behavior is depicted in
Fig.~\ref{figure19}(a), where we display the current-flux characteristics for
a half-filled ordered binary alloy ring with four electrons (two up and two 
down spin electrons). We see that for $U=2$ (dotted line) and $U=4$ (small
dashed line), current amplitudes are significantly larger than the 
non-interacting case, whereas for $U=10$ (dashed line) the current amplitude
becomes less than that from $U=0$ case. This enhancement takes place above 
quarter-filling i.e., when $N_e>N/2$ and can be easily understood as 
follows. As $N$ is even there are exactly $N/2$ number of sites with the 
lower site potential energy. If we do not take into account the 
electron-electron correlation, then above the quarter-filling it is 
\begin{figure}[ht]
{\centering\resizebox*{8cm}{5cm}{\includegraphics{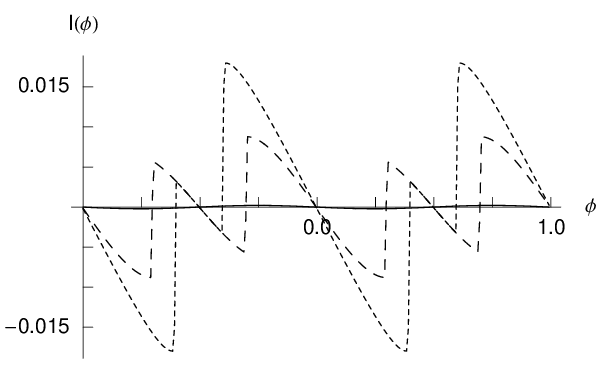}}\par}
\caption{Persistent current as a function of AB flux $\phi$ in a two 
($\uparrow,\downarrow$) electron incommensurate ring, where the solid, 
dotted and dashed curves correspond to $U=0$, $1$ and $3$, respectively.
The ring size $N$ is fixed at $30$.}
\label{figure21}
\end{figure}
preferred that some of these lower energy sites will be doubly occupied 
in the ground state. As we switch on the Hubbard correlation, the two 
electrons that are on the same site repel each other which results an 
enhancement in persistent current. But for large enough $U$, the hopping 
of the electrons is strongly suppressed by the interaction and the current 
amplitude gets reduced. On the other hand, at and below quarter-filling, 
no lower energy site will be doubly occupied in the ground state and hence 
there is no possibility of getting enhanced persistent current as a result
of Coulomb repulsion. In these systems we always get the reduction of 
persistent current with increasing the strength $U$. In 
Fig.~\ref{figure19}(b), we plot current-flux characteristics for a 
quarter-filled binary-alloy ring with $U=0$, $2$, $4$ and $6$, and it 
clearly noticed that persistent current is always suppressed due to this 
electronic correlation.

\subsubsection{Rings with incommensurate site potentials}

In this sub-section, we illustrate the behavior of persistent current in 
one-dimensional mesoscopic rings subject to quasi-periodic site potentials 
and study the effect of e-e interaction on persistent current. For a 
$N$-site ring with incommensurate site potentials, the TB Hamiltonian reads,
\begin{eqnarray}
H & = & \sum_{\sigma}\sum_{i=1}^{N}\epsilon \cos(i \lambda \pi)
c_{i,\sigma}^{\dagger} c_{i,\sigma} \nonumber \\
 & + & t\sum_{\sigma}\sum_{i=1}^{N}
\left(c_{i,\sigma}^{\dagger}c_{i+1,\sigma} e^{i\theta}+ 
c_{i+1,\sigma}^{\dagger}c_{i,\sigma} e^{-i\theta} \right) 
\nonumber \\
 & + & U\sum_{i=1}^{N} n_{i\uparrow}n_{i\downarrow}
\label{equ17}
\end{eqnarray}
where, $\lambda$ is an irrational number and as a typical example we 
take it the golden mean, $\frac{1+\sqrt 5}{2}$. For $\lambda=0$, we get 
back the pure ring with identical site potential $\epsilon$.

\vskip 0.2cm
\noindent
{\underline{Rings with two opposite spin electrons}:}
To investigate the precise role of electron-electron interaction on 
persistent current in presence of incommensurate site potentials let 
us first begin our discussion with a simplest possible system which is 
the case of a ring with two opposite spin (up and down) electrons. 
Figure~\ref{figure21} shows the current-flux characteristics of a $30$-site 
incommensurate ring, where the solid, dotted and dashed lines correspond 
to $U=0$, $1$ and $3$, respectively. The current gets reduced significantly
by the incommensurate site potentials in absence of any electron correlation. 
This is clearly observed from the solid line of Fig.~\ref{figure21}, where 
it almost coincides with the abscissa. This is due to the fact that, in 
presence of aperiodic site potentials the electronic energy eigenstates are 
critical~\cite{kohmoto,chakra1} which tends to localize the electrons,
and accordingly, the current amplitude gets decreased. But this situation
changes quite dramatically as we include electron-electron interaction.
From Fig.~\ref{figure21} it is clearly observed that the Hubbard correlation
considerably enhances persistent current for the low values of $U$ (dotted
curve). The reason is that the repulsive Coulomb interaction does not allow 
double occupancy of the sites in the ground state and also it opposes the 
confinement of electrons as a result of localization. Therefore, the
mobility of electrons increases as we introduce the electron-electron 
interaction and current gets enhanced. But such enhancement ceases to occur 
after certain values of $U$ due to the ring geometry, and the current then 
decreases as we increase the correlation strength $U$ further (dashed curve). 
Here we also observe that some strange kink-like structures appear in 
current-flux characteristics around $\phi=\pm 0.5$ for finite values of $U$ 
and inside these kinks persistent currents are independent of the Hubbard 
correlation strength. Therefore, it reveals that the kinks appear from 
the $U$-independent energy eigenstates of these rings. The explanation for 
the appearance of such kinks in persistent current has already been 
discussed in our previous sub-section for ordered rings. In these rings 
\begin{figure}[ht]
{\centering\resizebox*{8cm}{10cm}{\includegraphics{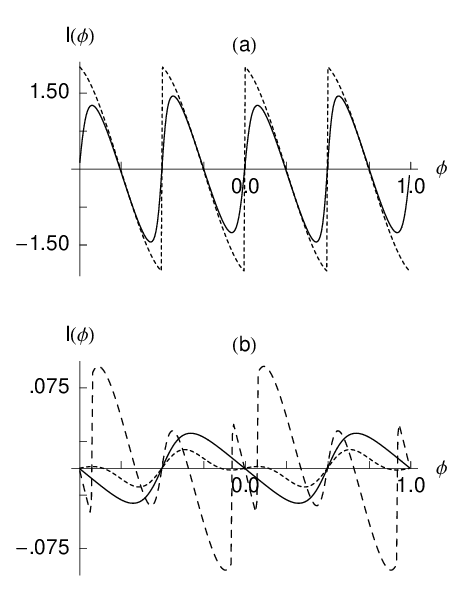}}\par}
\caption{Current-flux characteristics in a three ($\uparrow,\uparrow,
\downarrow$) electron incommensurate ring. (a) Half-filled ring ($N=3$)
with $U=4$, where the dotted and solid curves correspond to $\lambda=0$ 
and $\lambda \ne 0$, respectively. (b) Non-half-filled ($N=12$) ring with
$\lambda \ne 0$, where the solid, dotted and dashed curves represent 
$U=0$, $4$ and $50$, respectively.} 
\label{figure22}
\end{figure}
with two opposite spin electrons, the current shows kink-like structures 
as long as the interaction is included, but in presence of the incommensurate 
site potentials kinks appear above some critical value of the correlation 
strength depending on the system size and strength of randomness. For such 
two-electron incommensurate rings we get $\phi_0$ periodic currents.

\vskip 0.2cm
\noindent
{\underline{Rings with two up and one down spin electrons}:}
As representative examples of three-electron systems now we consider
incommensurate rings with two up and one down spin electrons.
In Fig.~\ref{figure22}(a) we present current-flux characteristics for an 
half-filled ring ($N=3$ and $N_e=3$) when the correlation strength $U$ is
set at $4$. The dotted curve of this spectrum corresponds to a pure ring 
($\lambda=0$) which exhibits discontinuous jumps at $\phi=0$, $\pm 0.5$ 
due to the crossing of energy levels. Quite interestingly we observe that, 
this pure half-filled three-electron system exhibits a perfect $\phi_0/2$ 
periodicity and we will see that this is a characteristic feature of pure 
half-filled rings with odd number of electrons. From the solid curve of 
this spectrum it is evident that, such $\phi_0/2$ periodicity no 
longer exists as we introduce the incommensurate site potentials and we get
back the usual $\phi_0$ flux-quantum periodicity. Moreover, in this case 
the current $I(\phi)$ becomes a continuous function with $\phi$ as the 
perturbation due to disorder lifts the ground state degeneracy at the 
\begin{figure}[ht]
{\centering\resizebox*{8cm}{9cm}{\includegraphics{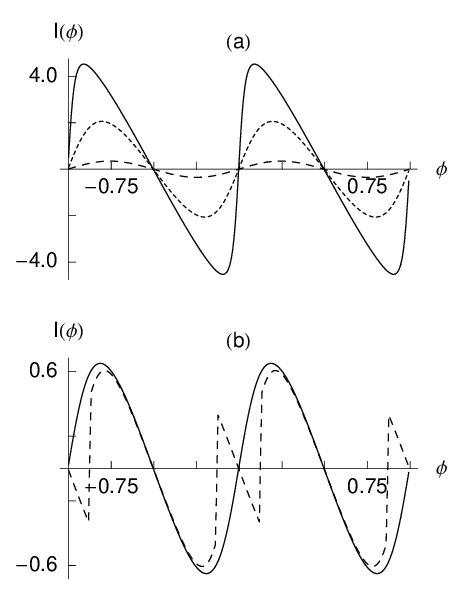}}\par}
\caption{Current-flux characteristics in a four ($\uparrow,\uparrow,
\downarrow,\downarrow$) electron incommensurate ring. (a) Half-filled 
($N=4$) band case, where the solid, dotted and dashed lines correspond 
to $U=0$, $4$ and $10$, respectively. (b) Non-half-filled ($N=8$) band 
case, where the solid and dashed lines are for $U=4$ and $16$, respectively.} 
\label{figure23}
\end{figure}
crossing points of the energy levels. The characteristic features of 
persistent current are quite different in the non-half-filled rings with 
two up and one down spin electrons. Figure~\ref{figure22}(b) gives the 
results for a $12$-site ring with incommensurate site potentials, where 
the solid, dotted and dashed lines represent $U=0$, $4$ and $50$,
respectively. The role of the Hubbard correlation on persistent current 
in presence of the incommensurate site potentials becomes evident from 
these characteristics curves. For low values of $U$, the $I$-$\phi$ curve 
resembles to that for the non-interacting case and the current does not
provide any discontinuity. But for large enough $U$, kink-like structures 
in persistent current are obtained as shown in Fig.~\ref{figure22}(b) by 
the dashed curve. In this case also the kinks are due to the $U$-independent 
eigenstates like the two electron systems, and as explained earlier the 
currents inside the kinks become independent of the correlation strength 
$U$. It is observed that persistent currents always have $\phi_0$ 
periodicity in the non-half-filled systems. Here we also notice that for 
the half-filled systems, current always decreases with the increase of 
$U$, while in the non-half-filled rings current gets significant 
enhancement due to interplay between the electron correlation and the 
incommensurate site potentials.

\vskip 0.2cm
\noindent
{\underline{Rings with two up and two down spin electrons}:}
Next we address the characteristic features of persistent current in 
four-electron systems with incommensurate site potentials and as 
illustrative examples we consider rings with two up and two down spin 
electrons. In Fig.~\ref{figure23}(a), we plot current-flux characteristics 
for an half-filled ring ($N=4$ and $N_e=4$), where the solid, dotted and 
dashed lines are for the cases with $U=0$, $4$ and $10$, respectively. From 
\begin{figure}[ht]
{\centering\resizebox*{8cm}{4.5cm}{\includegraphics{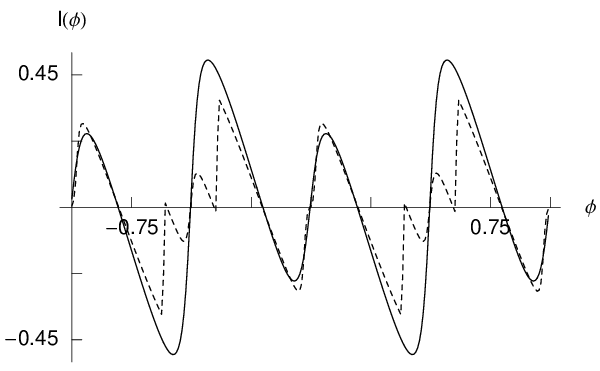}}\par}
\caption{Current-flux characteristics of a five ($\uparrow,\uparrow,
\uparrow,\downarrow,\downarrow$) electron incommensurate ring away from the
half-filled band case where the solid and dotted curves correspond to $U=18$ 
and $120$, respectively. The ring size $N$ is fixed at $7$.}
\label{figure24}
\end{figure}
these curves it is clearly observed that the current amplitude gradually 
decreases with the increase of the correlation strength $U$. This reveals 
that in the half-filled case, electronic mobility gradually decreases with 
the increase of $U$, and we see that for large enough $U$, the system goes 
to an insulating phase. This kind of behavior holds true for any half-filled 
system, because at large enough $U$ every site will be occupied by a single 
electron and the hopping of the electrons will not be favored due to strong
electron-electron repulsion. Figure~\ref{figure23}(b) displays the 
current-flux characteristics for a non-half-filled four-electron system 
with the aperiodic Harper potential. The solid and dotted curves are 
the $I$-$\phi$ curves for a $8$-site ring with $U=4$ and $U=10$, 
respectively. This figure depicts that for low values of $U$, persistent 
current $I(\phi)$ has no discontinuity but kinks appear in the $I$-$\phi$ 
curve at 
large value of $U$. These kinks appear at sufficiently large value of $U$ 
due to additional crossing of the ground state energy levels as we vary 
$\phi$. It is important to note that, in the present case kinks appear due 
to the $U$-dependent states and not from the $U$-independent states as in 
the previous two- and three-electron cases. Both for the half-filled or 
non-half-filled incommensurate rings with four electrons, we observe that 
persistent current always exhibits $\phi_0$ periodicity.

\vskip 0.2cm
\noindent
{\underline{Rings with three up and two down spin electrons}:}
Finally, we take five-electron aperiodic rings and evaluate persistent 
current in rings with three up and two down spin electrons. In a pure 
half-filled ring ($N=5, N_e=5$ and $\lambda=0$), we get $\phi_0/2$
periodic persistent current and we have already observed such period 
halving in other pure half-filled systems with odd number of electrons
({\em e.g.}, $N=3,~N_e=3$ and $\lambda=0$). Like the three-electron 
half-filled incommensurate rings, also in this case the $\phi_0$ periodicity 
of the persistent current is restored once we introduce the incommensurate 
site potentials. The current-flux characteristics for the non-half-filled 
five-electron rings with $N=7$ are shown in Fig.~\ref{figure24}. The solid 
and dotted curves correspond to $U=18$ and $120$, respectively. Similar to
non-half-filled three-electron system, here also kinks appear in persistent
current beyond a critical value of $U$. Another important observation is 
that for large $U$ ($U=120$), the maximum 
amplitude of the current remains finite. This is quite obvious since we  
consider the systems with $N>N_e$, where some sites become always empty so 
that electrons can hop to the empty sites which results the conducting phase. 
We also see that in these non-half-filled five-electron rings persistent 
currents always have the $\phi_0$ periodicity.

Thus the analysis of persistent current in ordered binary alloy and 
aperiodic Hubbard rings yields many interesting results due to interplay
between the electron-electron interaction and disorder in these systems.
The significant results are: (a) In absence of electron correlation, the
discontinuity in current-flux characteristics disappears due to disorder. 
This has been observed both in the ordered binary alloy rings and also in 
the aperiodic rings. (b) In pure rings with electron correlation, we 
observe both $\phi_0$ and $\phi_0/2$ periodicities in persistent currents. 
However, in the incommensurate and ordered binary alloy rings persistent 
currents always have the $\phi_0$ periodicity. (c) In ordered binary alloy 
rings, above the quarter-filling we get the enhancement of persistent current 
for small values of $U$ and the current eventually decreases when $U$ becomes 
large. On the other hand, at and below quarter-filling, persistent current 
always decreases with the increase of $U$. (d) An important finding is the 
appearance of kink-like structures in $I$-$\phi$ curves of the incommensurate 
rings only when we take into account the electron-electron interaction. Quite 
surprisingly we observe that, in some cases the currents inside the kinks are 
independent of the correlation strength $U$. 

\section{Enhancement of persistent current in one-channel rings and 
multi-channel cylinders}

Almost all the existing theories are basically based on the framework of 
nearest-neighbor TB model with either diagonal or off-diagonal disorder,
and it has been observed that the simple nearest-neighbor TB Hamiltonian 
cannot explain the observed enhancement of persistent current, even in 
presence of electron-electron interaction. In this sub-section, we will 
address the problem of the enhancement of persistent current in 
single-isolated-disordered mesoscopic one-channel rings and multi-channel 
cylinders, considering higher order hopping integrals in the Hamiltonian 
within a non-interacting electron picture, on the basis that the overlap 
of atomic orbitals between various neighboring sites are usually 
non-vanishing, and the higher order hopping integrals become quite 
significant~\cite{sanu4,sanu5,sanu6}. Physically, the higher order hopping 
integrals try to delocalize electrons even in one-dimension preserving 
their phase coherence and prevent the reduction of persistent current due 
to disorder. The fluctuations in persistent currents are also highly 
diminished due to the higher order hopping integrals. As a result, average 
amplitude of persistent current becomes comparable to $I_0$ and this is 
exactly what has been observed experimentally.

\subsection{One-channel mesoscopic rings}

\begin{figure}[ht]
{\centering\resizebox*{5.25cm}{3cm}{\includegraphics{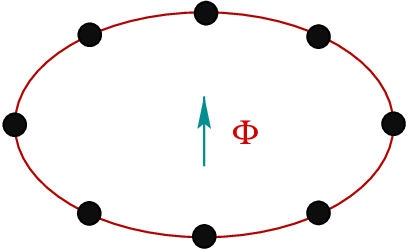}}\par}
\caption{(Color online). One-channel normal metal ring pierced by a magnetic 
flux $\phi$. The filled circles correspond to the positions of the lattice 
sites.}
\label{figring}
\end{figure}

We describe a $N$-site ring (Fig.~\ref{figring}) enclosing a magnetic 
flux $\phi$ (in units of the elementary flux quantum $\phi_0$) by the 
following TB Hamiltonian in the Wannier basis,
\begin{equation}
H=\sum_i \epsilon_i c_i^{\dagger} c_i + \sum_{i\ne j} t_{ij}\left(c_i^{\dagger}
c_j e^{i \theta_{ij}} + c_j^{\dagger} c_i e^{-i \theta_{ij}} \right)
\label{equ18}
\end{equation}
where, $\epsilon_i$ is the on-site energy and the phase factor $\theta_{ij}=
2\pi \phi\left(|i-j|\right)/N$. Here we take the hopping integral between any
two sites $i$ and $j$ through the expression $t_{ij}=t\exp\left[\alpha\left(1-
|i-j|\right)\right]$, where $t$ corresponds to the nearest-neighbor hopping 
(NNH) strength. Since the hopping integrals between far enough sites give 
negligible contributions, we consider only one higher order hopping 

integral in addition to the NNH integral which provides the hopping of an 
electron in the {\em next shortest path} between two sites. Therefore, in the
case of strictly one-channel rings the next possible shortest path becomes 
the twice of the lattice spacing.

\subsubsection{Impurity free rings}

Here we concentrate on the behavior of persistent current both for impurity 
free mesoscopic rings described with only NNH integral, and rings described 
\begin{figure}[ht]
{\centering\resizebox*{8cm}{7cm}{\includegraphics{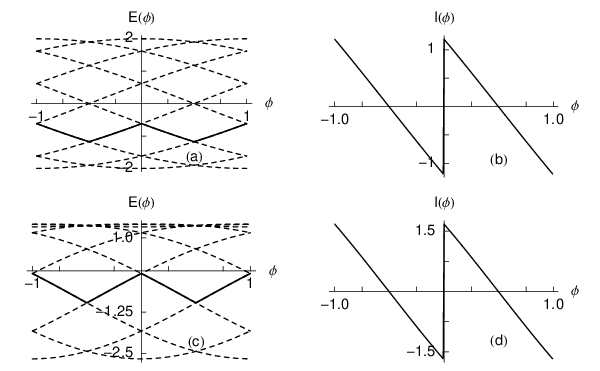}}\par}
\caption{Energy spectra and persistent currents of $10$-site ordered 
rings with four electrons ($N_e=4$), where (a) and (b) correspond to NNH 
model, while (c) and (d) correspond to SNH ($\alpha=1.1$) model.}
\label{figure26}
\end{figure}
with NNH integral in addition to the second-neighbor hopping (SNH) integral. 
In absence of any impurity, setting $\epsilon_i=0$ in Eq.~\ref{equ18}, the 
energy eigenvalue of the $n$th eigenstate can be expressed as,
\begin{equation}
E_n(\phi)=\sum_{p=1}^{p_0} 2 t \exp\left[\alpha\left(1-p \right) \right]
\cos\left[\frac{2 \pi p}{N}\left(n+\phi \right) \right]
\label{equ19}
\end{equation}
and the corresponding persistent current carried by this eigenstate becomes,
\begin{equation}
I_n(\phi) = \left(\frac{4 \pi t}{N}\right) \sum_{p=1}^{p_0} \exp\left[\alpha
\left(1-p \right) \right] \sin\left[\frac{2 \pi p}{N}\left(n+\phi \right) 
\right] \\
\label{equ20}
\end{equation}
where, $p$ is an integer. We take $p_0=1$ and $2$, respectively, for the 
rings with NNH and SNH integrals.
For large values of $\alpha$, the ring described with SNH integral 
eventually equivalent to the ring with only NNH integral. The contributions
from the SNH integral become much more appreciable only when we decrease the
value of $\alpha$. In such case, the energy spectrum and the persistent 
current get modified and these modifications provide some interesting new 
results, which can be available from the following analysis.

\begin{figure}[ht]
{\centering\resizebox*{8cm}{10cm}{\includegraphics{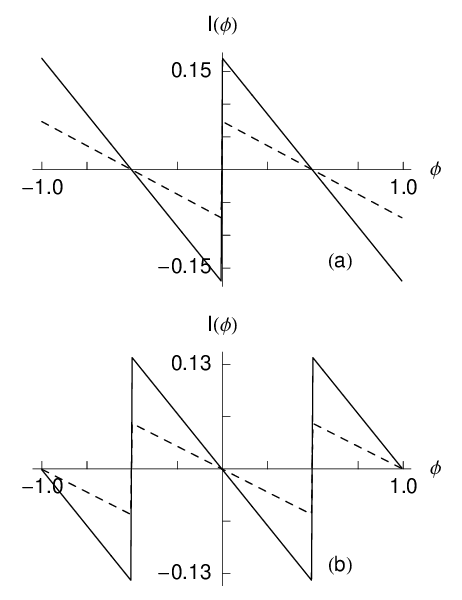}}\par}
\caption{Persistent current as a function of flux $\phi$ for some ordered 
rings with $N=100$ and $\alpha=0.9$, where (a) $N_e=20$ and (b) $N_e=15$. 
The dotted and solid curves correspond to the rings with NNH and SNH 
integrals, respectively.}
\label{figure27}
\end{figure}

To have a deeper insight to the problem, let us first describe the energy 
spectra and persistent currents of some small perfect rings ($N=10$) 
containing four electrons ($N_e=4$). The results are shown in 
Fig.~\ref{figure26}. The energy spectra for the rings described with NNH 
and SNH integrals are shown in Figs.~\ref{figure26}(a) and (c), respectively. 
Here the solid curves represent the variation of the Fermi level at $T=0$K 
with flux $\phi$. We see that the SNH integral lowers the energy levels, 
and most 
importantly, below the Fermi level the slopes of the $E(\phi)$ versus $\phi$ 
curves increases. As a result, the current gets increased in presence of SNH 
integrals and this enhancement of persistent current is clearly observed 
from Figs.~\ref{figure26}(b) and (d). In Fig.~\ref{figure26}, we have 
considered a $10$-site ring only for the sake of illustration and the 
results for a larger ring are presented in Fig.~\ref{figure27}.

In Fig.~\ref{figure27}, we plot current-flux characteristics for some
perfect rings with $N=100$ and $\alpha=0.9$. The dotted and solid curves
correspond to the variation of persistent current with flux $\phi$ for 
the rings described with NNH and SNH integrals, respectively. The enhancement 
of current amplitude due to the inclusion of SNH integral is clearly visible 
from Figs.~\ref{figure27}(a) and (b) by comparing the results plotted by the 
dotted and the solid curves. Figure~\ref{figure27}(a) shows that the current 
has sharp transitions at $\phi=0$ or $\pm n\phi_0$, while in 
Fig.~\ref{figure27}(b) the current shows transitions at $\phi=\pm n\phi_0/2$. 
These transitions are due to the degeneracy of energy eigenstates at these 
respective fields. For all the above models currents are always periodic in 
$\phi$ providing $\phi_0$ flux-quantum periodicity.

\subsubsection{Rings with impurity}

In order to understand the role of higher order hopping integral on 
persistent current in disordered mesoscopic rings, we first describe 
the energy spectra and persistent currents in small rings. The results 
\begin{figure}[ht]
{\centering\resizebox*{8cm}{8cm}{\includegraphics{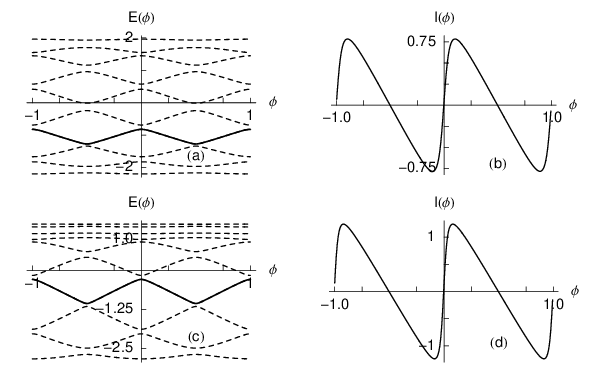}}\par}
\caption{Energy spectra and the persistent currents of $10$-site disordered 
($W=1$) rings with four electrons ($N_e=4$), where (a) and (b) correspond 
to NNH model, while (c) and (d) correspond to SNH ($\alpha=1.1$) model.}
\label{figure28}
\end{figure}
for a $10$-site disordered ring with $N_e=4$ are shown in 
Fig.~\ref{figure28}. To describe the system we use the tight-binding 
Hamiltonian as given in Eq.~\ref{equ18}, where the site energies 
($\epsilon_i$'s) are chosen randomly from a ``Box" distribution function
of width $W$. The energy spectra for the rings described with NNH and SNH 
integrals are plotted in Figs.~\ref{figure28}(a) and (c), respectively. 
In these figures the solid curves give the location of the Fermi level. 
Like the ordered cases, the SNH integral lowers the energy levels and below 
the Fermi level slopes of the $E(\phi)$ versus $\phi$ curves become much 
more than those for the NNH model. Thus even in the presence of impurity, 
we get the enhancement of persistent current due to the SNH integral, which 
is observed from the results presented in Figs.~\ref{figure28}(b) and (d), 
respectively.

\begin{figure}[ht]
{\centering\resizebox*{8cm}{8.5cm}{\includegraphics{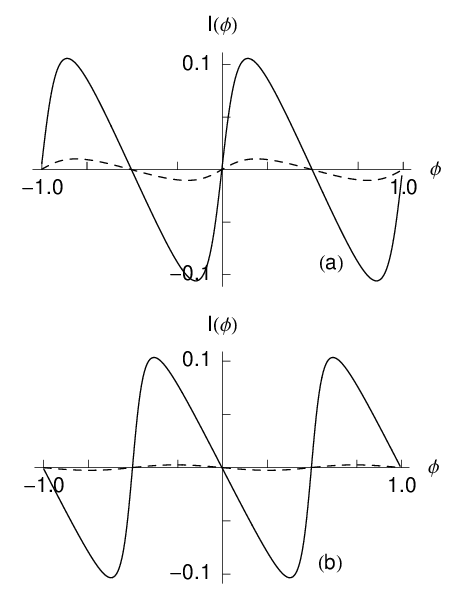}}\par}
\caption{Persistent current as a function of flux $\phi$ for some disordered 
rings with $N=100$, $\alpha=0.9$ and $W=1$, where (a) $N_e=20$ and 
(b) $N_e=15$. The dotted and solid curves correspond to the rings with 
NNH and SNH integrals, respectively.}
\label{figure29}
\end{figure}

In Fig.~\ref{figure29}, we plot the current-flux characteristics for some
larger disordered rings considering $N=100$, $\alpha=0.9$ and the disorder
strength $W=1$. The dotted and solid lines correspond to the rings described 
with NNH and SNH integrals, respectively. The results for the even number 
of electrons ($N_e=20$) are shown in Fig.~\ref{figure29}(a), while 
Fig.~\ref{figure29}(b) represents the currents for the odd $N_e$ ($N_e=15$). 
The currents are computed for some typical disordered configurations of the 
ring, and in fact we observe that the qualitative nature of persistent 
current does not depend on the specific realization of the disordered 
configurations. Figure~\ref{figure29} shows that persistent current
for the disordered rings is always periodic in $\phi$ with $\phi_0$ 
flux-quantum periodicity. In the presence of impurity, current becomes a 
continuous function of flux $\phi$ which is clearly visible from this figure 
(Fig.~\ref{figure29}). For the perfect rings, sharp transitions in
persistent current at the points $\phi=0$ or $\pm n\phi_0$ with even $N_e$ 
and at $\phi=\pm n\phi_0/2$ with odd $N_e$ appear due to the degeneracy of 
the ground state energy at these respective field points. Now as the 
impurities are introduced, all the degeneracies get lifted and the current 
exhibits a continuous variation with respect to the flux $\phi$. At these 
degenerate points, the ground state energy passes through an extrema which 
in turn gives the zero persistent current as shown in Fig.~\ref{figure29}.
It is clear from Fig.~\ref{figure29} that, the second-neighbor hopping (SNH) 
integral plays a significant role to enhance the amplitude of persistent
current in disordered rings. From Figs.~\ref{figure29}(a) and (b) we see 
that, the currents in disordered rings with only NNH integrals (dotted 
curves) are vanishingly small compared to those as observed in the impurity 
free rings with NNH integrals (dotted curves in Figs.~\ref{figure27}(a) and
(b)). On the other hand, Fig.~\ref{figure29} emphasizes that the currents 
in the disordered rings with higher order hopping integral are of the 
same order of magnitude as those for the ordered rings.

\begin{figure}[ht]
{\centering\resizebox*{8cm}{6.5cm}{\includegraphics{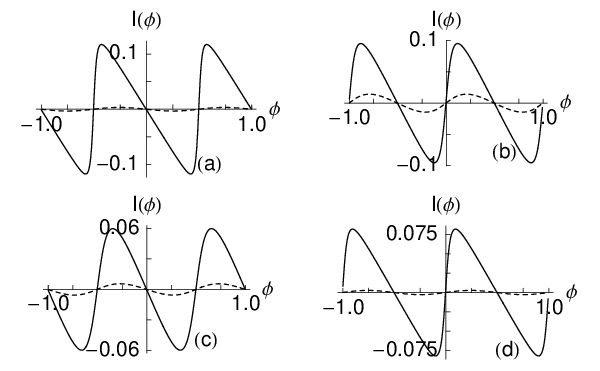}}\par}
\caption{Current-flux characteristics for some disordered rings with higher 
electron concentrations and we set $W=1$ and $\alpha=1.1$, where (a) $N=125$, 
$N_e=45$; (b) $N=125$, $N_e=40$; (c) $N=150$, $N_e=55$ and (d) $N=150$, 
$N_e=60$.}
\label{figure30}
\end{figure}

In Fig.~\ref{figure30}, we plot persistent currents for some disordered 
rings with higher electron concentrations, and study the cases with $N=$ 
even or odd and $N_e=$ even or odd. The dotted and solid curves 
respectively corresponds to the NNH and SNH models. It is observed that 
the evenness or the oddness of $N$ and $N_e$ do not play any important role 
on persistent current, but we will see later that the diamagnetic or the
paramagnetic sign of the current crucially depends on the evenness or 
oddness of $N_e$.

Physically, the higher order hopping integrals try to delocalize electrons 
preserving their phase coherence and prevent the reduction of the current 
due to disorder. In disordered 
rings with only NNH integrals, the enormous reduction of the current 
amplitudes are basically due to localization of the energy eigenstates. 
When we add higher order hopping integrals, it is most likely that the 
localization length increases and may become comparable to the length of 
the ring, and we get the enhancement of the persistent current.

\subsection{Multi-channel mesoscopic cylinders}

\begin{figure}[ht]
{\centering\resizebox*{7cm}{4.5cm}{\includegraphics{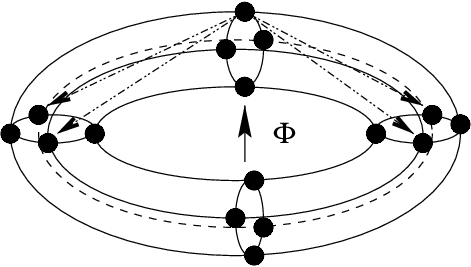}}\par}
\caption{Schematic view of a multi-channel mesoscopic cylinder threaded by
a magnetic flux $\phi$. The filled circles correspond to the positions of 
the lattice sites.}
\label{figure31}
\end{figure}

So far we have confined our discussions only to one-dimensional systems which
do not really correspond to the experimental situations. Enhancement of the
persistent current has been observed even in single-isolated diffusive
(disordered) metallic rings. But diffusion is not possible strictly in 
one-dimension and it becomes necessary to consider finite width of the 
samples~\cite{sanu6}. The simplest way of doing this is to consider a 
cylindrical mesoscopic ring threaded by a magnetic flux $\phi$. 
A schematic representation of the system is given in Fig.~\ref{figure31}. 
Assuming that the lattice spacing both in the longitudinal and transverse 
directions are identical (i.e., surface of the cylinder forms a square 
lattice), we can describe the system by the TB Hamiltonian as,
\begin{equation}
H = \sum_x \epsilon_x c_x^{\dagger}c_x + \sum_{<xx^{\prime}>}\left[
t_{x x^{\prime}} e^{i\theta_{x x^{\prime}}} c_x^{\dagger} c_{x^{\prime}} 
+ t_{x x^{\prime}} e^{-i\theta_{x x^{\prime}}} c_{x^{\prime}}^{\dagger}c_x 
\right] 
\label{equ21}
\end{equation}
where $\epsilon_x$ is the site energy of the lattice point $x$ of coordinate,
say, ($i$, $j$). $t_{xx^{\prime}}$ is the hopping integral between the
lattice points $x$ and $x^{\prime}$ and $\theta_{x x^{\prime}}$ is the
phase factor acquired by the electron due to this hopping in presence of 
magnetic flux $\phi$. Let us now investigate the role of just the 
second-neighbor hopping integral on persistent current, and neglect the 
effects of all higher order hopping integrals. Let $t$ denotes the 
nearest-neighbor hopping integral and the second-neighbor hopping
integral (across the diagonal of the square) is taken to have the 
exponential form: $t \exp(-\alpha)$, where $\alpha$ is the decay constant.

In absence of any impurity, setting $\epsilon_x=0$, the energy eigenvalue
of the $n$th eigenstate becomes,
\begin{eqnarray}
E_n(\phi) & = & 2 t\cos\left[\frac{2\pi}{N}(n+\phi)\right] 
+ 4 t e^{-\alpha}\cos\left[\frac{2\pi}{N}(n+\phi)\right] \nonumber \\
& & \times \cos\left[\frac{2\pi m}{M}\right] 
+ 2 t \cos\left[\frac{2\pi m}{M}\right]
\label{equ22}
\end{eqnarray}
and the persistent current carried by this eigenstate is,
\begin{eqnarray}
I_n(\phi) & = & \left(\frac{4\pi t}{N}\right)\sin\left[\frac{2\pi}{N}(n+\phi)
\right] 
+\left(\frac{8\pi t}{N}\right) e^{-\alpha} \nonumber \\
& &\times \sin\left[\frac{2\pi}{N}(n+\phi)\right]
\cos\left[\frac{2\pi m}{M}\right]
\label{equ23}
\end{eqnarray}
where, $n$ and $m$ are two integers bounded within the range 
$-\lfloor N/2 \rfloor \leq n < \lfloor N/2 \rfloor$ and
$-\lfloor M/2 \rfloor \leq m < \lfloor M/2 \rfloor$, respectively, where
$\lfloor \ldots \rfloor$ denotes the integral part. Here $M$ and $N$ are
the number of sites along the longitudinal and transverse directions 
of the cylinder, respectively.

\begin{figure}[ht]
{\centering\resizebox*{8cm}{9cm}{\includegraphics{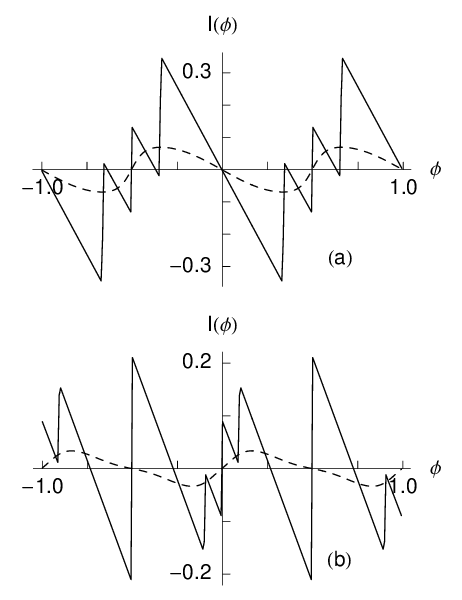}}\par}
\caption{Persistent current as a function of flux $\phi$ for a typical 
multi-channel mesoscopic cylinder with $N=50$ and $M=4$ described by NNH 
integral only, where (a) $N_e=45$ and (b) $N_e=40$. The solid and dotted 
lines are for the perfect ($W=0$) and disordered ($W=1$) cylinders,
respectively.}
\label{figure32}
\end{figure}

Let us first describe the behavior of persistent current in a multi-channel 
cylinder using the nearest-neighbor tight-binding Hamiltonian. The results
are shown in Fig.~\ref{figure32}, where (a) and (b) correspond to $N_e=45$
and $40$, respectively. Here we set $N=50$ and $M=4$. 
The solid curves describe the currents in absence of any impurity, while 
the dotted lines are for the disordered case and to introduce impurities
$\epsilon_x$'s are taken as random variables with uniform ``Box" 
distribution of width $W$. The persistent current for the perfect cylinder 
(solid curves) has many discontinuities within each $\phi_0$ flux-quantum 
period. These discontinuities are due to the existence of degenerate energy 
levels at certain magnetic flux, and these degeneracies get lifted as long
as impurities are included. The current for the disordered system provides 
a continuous variation with $\phi$ as shown by the dotted lines in 
Figs.~\ref{figure32}(a) and (b). It is observed that, even in 
multi-channel cylindrical systems the nearest-neighbor TB model gives 
orders of magnitude reduction of persistent currents compared to the 
results for the ballistic case.

\begin{figure}[ht]
{\centering\resizebox*{8cm}{9cm}{\includegraphics{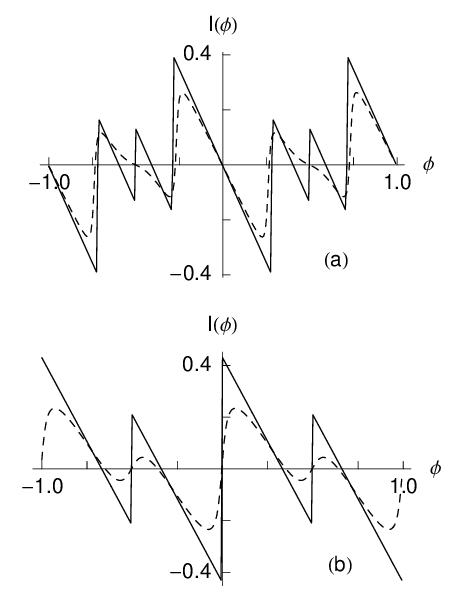}}\par}
\caption{Persistent current as a function of flux $\phi$ for a multi-channel 
mesoscopic cylinder described with both NNH and SNH ($\alpha=1.0$) integrals, 
where (a) $N_e=45$ and (b) $N_e=40$. The solid and dotted curves correspond 
to the perfect ($W=0$) and dirty ($W=1$) cylinders, respectively. Here we 
fix $N=50$ and $M=4$.}
\label{figure33}
\end{figure}

The behavior of persistent current for the disordered mesoscopic cylinder 
changes drastically as we switch on the second-neighbor hopping 
integrals. In Fig.~\ref{figure33}, we plot the current-flux characteristics 
for a multi-channel cylinder in presence of the second-neighbor hopping 
integral ($\alpha=1$) considering $M=50$ and $N=4$. The results shown in 
Figs.~\ref{figure33}(a) and (b) are respectively for the cylinders with 
$N_e=45$ and $40$, where the solid and dotted lines correspond to the 
identical meaning as in Fig.~\ref{figure32}. From the spectra we see that 
the current amplitudes in the disordered cylinder (dotted curves) are 
comparable to that of the perfect systems (solid curves). This is due to the 
fact that higher order hopping integrals try to delocalize the electrons, 
and accordingly, the current amplitudes get enhanced even by an order of
magnitude in comparison with the estimates of current amplitudes in 
disordered cylinders using the nearest-neighbor TB Hamiltonian. This study 
reveals that for both the mesoscopic one-channel rings and the multi-channel 
cylinders the higher order hopping integrals play a very significant role 
in the enhancement of persistent current amplitude in disordered systems.

\section{Low-field magnetic response on persistent current}

The diamagnetic or paramagnetic sign of low-field persistent currents also 
becomes a controversial issue due to discrepancy between theory and 
experiment. From the theoretical calculations, Cheung 
{\em et al.}~\cite{cheu2} predicted that the sign of persistent current 
is random depending on total number of electrons, $N_e$, in the system 
and on specific realization of the disordered configurations of the ring. 
Both the diamagnetic and paramagnetic responses were also observed 
theoretically in mesoscopic Hubbard rings by Yu and Fowler~\cite{yu}. They 
showed that the rings with odd $N_e$ exhibit paramagnetic response, while 
those with even $N_e$ have diamagnetic response in the limit 
$\phi \rightarrow 0$. In an experiment on $10^7$ isolated mesoscopic Cu 
rings, Levy {\em et al.}~\cite{levy} had reported diamagnetic response for 
the low-field currents, while with Ag rings Chandrasekhar 
{\em et al.}~\cite{chand} got the paramagnetic phase. In a recent experiment, 
Jariwala {\em et al.}~\cite{jari} have got diamagnetic persistent currents 
with both integer and half-integer flux-quantum periodicities in an array 
of $30$-diffusive mesoscopic gold rings. The diamagnetic sign of persistent
currents in the vicinity of zero magnetic field were also found in an 
experiment~\cite{deb} on $10^5$ disconnected Ag ring. The sign of the 
low-field current is a priori not consistent with the theoretical 
predictions. In this sub-section, we will study the nature of low-field 
magnetic response by calculating magnetic susceptibility of mesoscopic 
one-channel rings and multi-channel cylinders through some exact 
calculations.

The magnetic susceptibility of a $N$-site AB ring can be obtained 
from the general expression~\cite{sanu1},
\begin{equation}
\chi(\phi)=\frac{N^3}{16\pi^2}\left[\frac{\partial I(\phi)}{\partial \phi}
\right].
\label{equ24}
\end{equation}
Calculating the sign of $\chi(\phi)$, one can predict whether the current 
is paramagnetic or diamagnetic. Here we focus our attention on the systems 
either with fixed number of electrons $N_e$ or with fixed chemical 
potential $\mu$.

\subsection{One-channel mesoscopic rings}

\begin{figure}[ht]
{\centering \resizebox*{8cm}{9cm}{\includegraphics{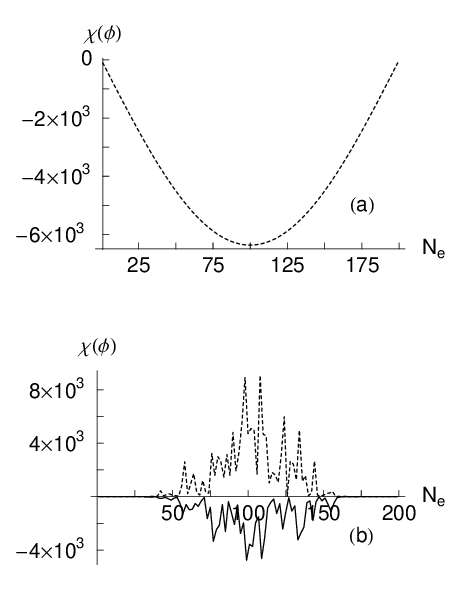}}\par}
\caption{Low-field magnetic susceptibility as a function of $N_e$ for
(a) perfect ring ($W=0$) and (b) disordered ($W=1$) ring. The solid and 
dotted lines in (b) correspond to the results for odd and even $N_e$, 
respectively. Here we set $N=200$.}
\label{figure35}
\end{figure}

Let us first analyze the behavior of low-field magnetic susceptibility in 
an impurity-free one-channel mesoscopic ring described with fixed $N_e$. 
Figure~\ref{figure35}(a) shows the variation of $\chi(\phi)$ as a function 
of $N_e$ for a perfect ring with $N=200$ in the limit $\phi \rightarrow 0$. 
It is noticed that, both for the even and odd $N_e$, current has only the 
diamagnetic sign. This diamagnetic sign of the low-field currents can be 
easily understood from the slope of current-flux curve of one-channel 
impurity-free rings (see the curves given in Fig.~\ref{figure3}). From 
these curves we observe that the current always exhibits negative slope 
at low-fields. Therefore, it can be predicted that for perfect one-channel 
rings current provides only the diamagnetic sign near zero-field limit, 
irrespective of the total number of electrons $N_e$ i.e., whether the 
rings contain odd or even $N_e$.

The effects of disorder on low-field currents are quite interesting, and
our results show that the sign of the currents, even in presence of 
disorder, can be mentioned without any ambiguity both for rings with odd 
and even $N_e$. In Fig.~\ref{figure35}(b), we plot $\chi(\phi)$ as a function 
of $N_e$ for the disordered case. Here we set $N=200$ and $W=1$. The solid 
and dotted lines correspond to the results for odd and even $N_e$, 
respectively. These curves show that the rings with odd $N_e$ exhibit only 
the diamagnetic sign for the low-field currents, while for even $N_e$ the 
low-field currents always have the paramagnetic sign. Physically, the 
disorder lifts all the degeneracies of the energy levels those were observed 
in a perfect ring, and as a result the sharp discontinuities of the 
$I$-$\phi$ curves (see the curves of Fig.~\ref{figure3}) disappear. It may 
be noted that the slopes of the $I$-$\phi$ curves for even and odd $N_e$ 
always have opposite signs near zero magnetic field (see Fig.~\ref{figure6}). 
Thus for one-dimensional disordered rings with fixed number of electrons, 
the sign of low-field current is independent of the specific realization of 
disordered configurations and depends only on the oddness or evenness of 
$N_e$.

\vskip 0.5cm
\begin{center}
{\bf Effect of temperature}
\end{center}
\vskip 0.2cm

At non-zero temperature, we notice an interesting behavior of low-field
magnetic susceptibility in mesoscopic rings. Let us confine ourselves to 
the systems with even number of electrons. At any finite temperature, 
magnetic response of these systems are always paramagnetic both for perfect 
and disordered rings in the zero field limit. For a given system, this 
paramagnetism is observed over a certain range of $\phi$ close to $\phi=0$, 
say, in the domain $\phi_0/4 \leq \phi \leq \phi_0/4$. Quite interestingly, 
we observe that, at finite temperatures the magnetic response of this 
particular system becomes diamagnetic beyond a critical field $\phi_c(T)$, 
even though $|\phi_c(T)| < \phi_0/4$.

\begin{figure}[ht]
{\centering\resizebox*{8cm}{5cm}{\includegraphics{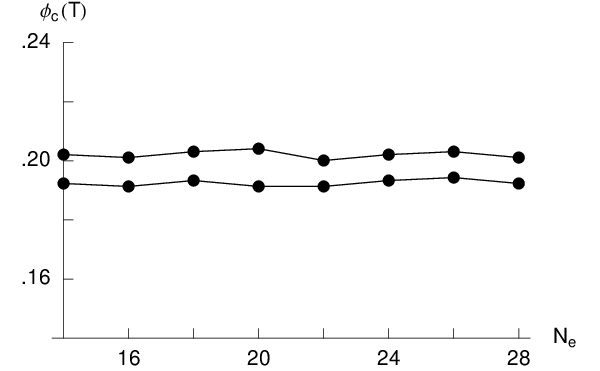}}\par}
\caption{$\phi_c(T)$ versus $N_e$ ($N_e$=even) curves for perfect rings with
size $N=45$.}
\label{figure36}
\end{figure}

\begin{figure}[ht]
{\centering \resizebox*{8cm}{5cm}{\includegraphics{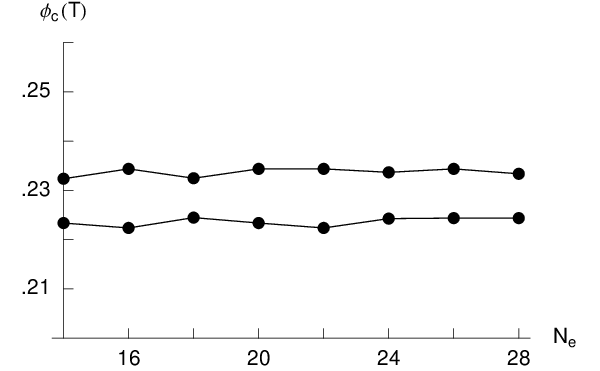}}\par}
\caption{$\phi_c(T)$ versus $N_e$ ($N_e$=even) curves for disordered ($W=1$)
rings with size $N=40$.}
\label{figure37}
\end{figure}

In Fig.~\ref{figure36}, we show the variation of the critical field 
$\phi_c(T)$ as a function of even $N_e$ for a perfect $45$-site one-channel 
ring. The curve with higher values of $\phi_c(T)$
corresponds to the temperature $T/T^{\star}=1.0$, while the other curve
corresponds to $T/T^{\star}=0.5$. Figure~\ref{figure37}, on the other hand, 
represents the behavior of $\phi_c(T)$ for a disordered sample (with $W=1$) 
at the same temperatures, $T/T^{\star}=1.0$ (upper curve) and 
$T/T^{\star}=0.5$ (lower curve). From these spectra (Figs.~\ref{figure36} 
and \ref{figure37}) it is clear that the critical value of $\phi$, where 
the transition from the paramagnetic to diamagnetic phase takes place, 
increases with the increase of the temperature. Thus we see that, both for 
the perfect and disordered rings with even number of electrons there exists 
a critical value of magnetic flux $\phi_c(T)$, beyond which the magnetic 
response of the low-field currents exhibits a transition from the 
paramagnetic to the diamagnetic one.

The situation is quite different even at zero temperature when we describe
the system by constant chemical potential instead of fixed $N_e$. It may be 
noted that, only for some particular values of $\mu$ the system will have 
a fixed number of electrons, and for these values of $\mu$ the sign of 
low-field currents can be predicted according to the above prescriptions. 
While, for all other choices of $\mu$, total number of electrons varies even 
for a slight change in magnetic flux $\phi$ in the neighborhood of zero flux. 
Hence, it is not possible to predict the
sign of the low-field currents precisely, even in absence of any impurity 
in the system. Thus the sign of low-field currents strongly depends on 
the choice of $\mu$, strength of disorder and disordered configurations.

\subsection{Multi-channel mesoscopic cylinders}

We have also studied the low-field magnetic response in mesoscopic 
rings of finite widths~\cite{sanu1}. Our study reveals that, for such 
systems it is not possible to predict precisely the sign of the low-field 
currents even for impurity-free cases with fixed number of electrons. 
So we can conclude that, in diffusive multi-channel mesoscopic rings sign 
of the low-field currents is a highly unpredictable quantity as it can be 
easily affected by total number of electrons $N_e$, chemical potential $\mu$, 
magnetic flux $\phi$, strength of disorder $W$, realizations of disordered 
configurations, etc. This is exactly the same picture what has been observed 
experimentally regarding the sign of the low-field currents.

\section{Concluding Remarks}

In the present review we have demonstrated quantum transport properties 
in different types of closed loop systems. These are one-channel
rings, multi-channel cylinders, etc.

At the beginning of this review (Section I), we have described very 
briefly some of the spectacular effects those appear in mesoscopic 
systems as a consequence of the quantum phase coherence of electronic 
wave functions. One of the most remarkable consequences of the quantum phase 
coherence is the appearance of AB oscillations in normal metal mesoscopic 
rings. Some other mesoscopic phenomena that were observed in mesoscopic 
systems are the integer and fractional quantum Hall effects, conductance 
fluctuations and its quantization, etc. In this review, we have first 
concentrated on the spectacular mesoscopic phenomenon where a non-decaying 
current, the so-called persistent current, circulates in a small metallic 
loop threaded by a slowly varying magnetic flux. To understand the behavior 
of experimental results on persistent current, one has to focus attention 
on the interplay of quantum phase coherence, electron-electron correlation 
and disorder. This is a highly challenging problem and here we have tried
to address this problem.

The characteristic features of persistent current in non-interacting 
single-channel rings and multi-channel cylinders have been presented in 
Section II showing its dependence on total number of electrons $N_e$, 
chemical potential $\mu$, randomness and total number of channels. All 
the calculations have been performed only at absolute zero temperature. 
In perfect one-channel rings with fixed $N_e$, persistent current shows 
saw-tooth like behavior as a function of magnetic flux $\phi$ with sharp 
discontinuities at $\phi=\pm n\phi_0/2$ or $\pm n \phi_0$ depending on 
whether the ring contains odd or even $N_e$. On the other hand, some 
additional kinks may appear in persistent currents for one-channel perfect 
rings described with fixed $\mu$. The situation is somewhat different for 
the multi-channel perfect cylinders. In such cylindrical rings, kinks 
appear in persistent currents for both the cases with fixed $N_e$ or 
fixed $\mu$. 

In Section III, we have explored the effects of e-e correlation 
and disorder on persistent current in single-channel rings. We have used the 
TB Hubbard model and determine persistent current by exact numerical 
diagonaliztion of the Hamiltonian. First, we have studied the behavior of 
persistent current in some perfect small with few number of electrons. We 
have found many interesting results those are: the appearance of kinks in 
persistent current due to electron-electron interaction, existence of both 
$\phi_0/2$ and $\phi_0$ flux-quantum periodicities in persistent current, 
disappearance of the singular behavior of persistent current in the 
half-filled rings with even number of electrons, existence of
$U$-independent energy eigenstates, appearance of both the metallic and 
insulating phases, etc. The discontinuities in persistent current 
at non-integer values of $\phi_0$ due to the electron correlation have 
also been observed, which crucially depend on the filling of the ring and 
also on the parity of the number of electrons. Next, we have investigated 
the effects of electron-electron correlation on persistent current in 
ordered binary alloy rings and aperiodic rings. The main results are: 
(a) In absence of electron correlation both for the ordered binary alloy 
and aperiodic rings the discontinuity in the $I$-$\phi$ curves disappears. 
(b) The persistent currents exhibit only $\phi_0$ flux-quantum periodicity. 
(c) In ordered binary alloy rings with more than quarter-filled, we observe 
enhancement of persistent current for small values of $U$, but it eventually 
decreases when $U$ becomes very large. On the other hand, at and below 
quarter-filling current always decreases with the strength of $U$. 

Though we have noticed some enhancement of current amplitude in disordered
rings due to Hubbard correlation, but still the amplitude is 
orders of magnitude smaller than the experimental estimates. In order to 
explain the enhancement of current amplitude, in Section IV we have 
calculated persistent currents in one-channel rings and multi-channel 
cylinders by inserting higher order hopping integrals together with 
nearest-neighbor hopping, within a non-interacting electron picture. 
The inclusion of the higher order hopping integrals is based on the fact 
that the overlap of atomic orbitals between various neighboring sites are 
usually non-vanishing, and these higher order hopping integrals try to 
delocalize electrons and prevent reduction of persistent current in presence
of disorder. It has also been observed that the fluctuations of persistent 
currents are also significantly diminished due to the higher order hopping 
integrals and the results are comparable to the experimental values.

The diamagnetic or the paramagnetic sign of the low-field currents is a 
controversial issue due to the discrepancy between theory and experiment. 
At the end (Section V) of this review, we have examined the behavior of
low-field magnetic response of persistent currents by calculating magnetic 
susceptibility in the limit $\phi \rightarrow 0$. In perfect one-channel 
rings, low-field current exhibits only the diamagnetic sign irrespective 
of the parity of the total number of electrons $N_e$ i.e, whether $N_e$ is 
odd or even, while in disordered rings currents have the diamagnetic or the 
paramagnetic nature depending on whether the rings contain odd or even $N_e$. 
The important point is that, for disordered one-channel rings with fixed 
$N_e$ the sign of the low-field currents is completely independent of 
the specific realization of the disordered configurations. In this context
we have also studied the effect of finite temperature and observed that 
both for the perfect and disordered rings containing even number of 
electrons, there exits a critical value of magnetic flux $\phi_c(T)$ beyond 
which the magnetic response of the low-field currents makes a transition 
from the paramagnetic to diamagnetic phase. But in disordered rings described
with fixed chemical potential $\mu$, the sign of low-field currents cannot 
be predicted since it strongly depends on the choices of $\mu$. Finally, in 
the case of multi-channel systems we have noticed that sign of these currents 
cannot be predicted exactly, even in the perfect case with fixed $N_e$ 
as it significantly depends on the choice of $N_e$, $\mu$, number of 
channels, disordered configurations, etc.

\vskip 0.2cm
\noindent
{\bf Future directions and opportunities:}
Although the studies involving persistent currents in mesoscopic rings and 
cylinders have already generated a wealth of literature there is still need 
to look deeper into the problems both from the point of view of fundamental
physics and to resolve a few issues that have not yet been answered in
an uncontroversial manner. For example, it may be extremely interesting to 
study thermal signatures of persistent currents in metallic rings. The 
recent progress in the research on persistent currents in metallic rings
suggests that by measuring heat capacity persistent current can be 
detected~\cite{therm1}, and this approach is completely different from the 
conventional methods i.e., by measuring the ring's magnetic moment using 
a SQUID magnetometer~\cite{levy,jari,chand,blu} or connecting the ring to 
a superconducting microresonator~\cite{deb}, or by using a sophisticated 
micromechanical detector~\cite{fvo}. The combined effect of 
electron-electron interaction and spin-orbit interaction on persistent
current is also an interesting topic which should be carefully examined.

\end{document}